\documentclass[final,3p,times,authoryear]{elsarticle}
\usepackage{amsmath}
\usepackage{multirow}
\usepackage{enumerate}
\usepackage{graphicx}
\usepackage{csvsimple}
\usepackage{algorithm}
\usepackage{algcompatible}
\usepackage{multicol}
\usepackage{url}
\usepackage{float}
\usepackage{cite}

\usepackage{lscape}
\usepackage{longtable}

\usepackage{xcolor}
\usepackage{fancyhdr}

\DeclareMathOperator*{\minimize}{\textbf{minimize}}

\usepackage{afterpage}

\usepackage{amssymb}

\journal{Journal of Network and Computer Applications}

\begin{document}

\begin{frontmatter}



\title{An Algorithm for Network and Data-aware Placement of Multi-Tier Applications in Cloud Data Centers}


\author[MU,UM]{Md Hasanul Ferdaus\corref{cor1}}
\ead{md.ferdaus@monash.edu}
\cortext[cor1]{Corresponding author}

\author[FedUni]{Manzur Murshed}
\ead{manzur.murshed@federation.edu}

\author[UM]{Rodrigo N. Calheiros}
\ead{rnc@unimelb.edu.au}

\author[UM]{Rajkumar Buyya}
\ead{rbuyya@unimelb.edu.au}

\address[MU]{Faculty of Information Technology, 25 Exhibition Walk, Clayton campus, Monash University, VIC 3800, Australia}

\address[FedUni]{Faculty of Science and Technology, Federation University Australia, Northways Road, Churchill, VIC 3842, Australia}

\address[UM]{Cloud Computing and Distributed Systems (CLOUDS) Laboratory, Department of Computing and Information Systems, Building 168, The University of Melbourne, Parkville, VIC 3053, Australia}

\begin{abstract}
Today's Cloud applications are dominated by composite applications comprising multiple computing and data components with strong communication correlations among them. Although Cloud providers are deploying large number of computing and storage devices to address the ever increasing demand for computing and storage resources, network resource demands are emerging as one of the key areas of performance bottleneck. This paper addresses network-aware placement of virtual components (computing and data) of multi-tier applications in data centers and formally defines the placement as an optimization problem. The simultaneous placement of Virtual Machines and data blocks aims at reducing the network overhead of the data center network infrastructure. A greedy heuristic is proposed for the on-demand application components placement that localizes network traffic in the data center interconnect. Such optimization helps reducing communication overhead in upper layer network switches that will eventually reduce the overall traffic volume across the data center. This, in turn, will help reducing packet transmission delay, increasing network performance, and minimizing the energy consumption of network components. Experimental results demonstrate performance superiority of the proposed algorithm over other approaches where it outperforms the state-of-the-art network-aware application placement algorithm across all performance metrics by reducing the average network cost up to 67\% and network usage at core switches up to 84\%, as well as increasing the average number of application deployments up to 18\%.
\end{abstract}


\end{frontmatter}


\section{Introduction}
\label{chap4-sec-intro}

\begin{figure}[!t]
\centering
\includegraphics[scale=.75, trim=0cm 18.3cm 1cm 2cm]{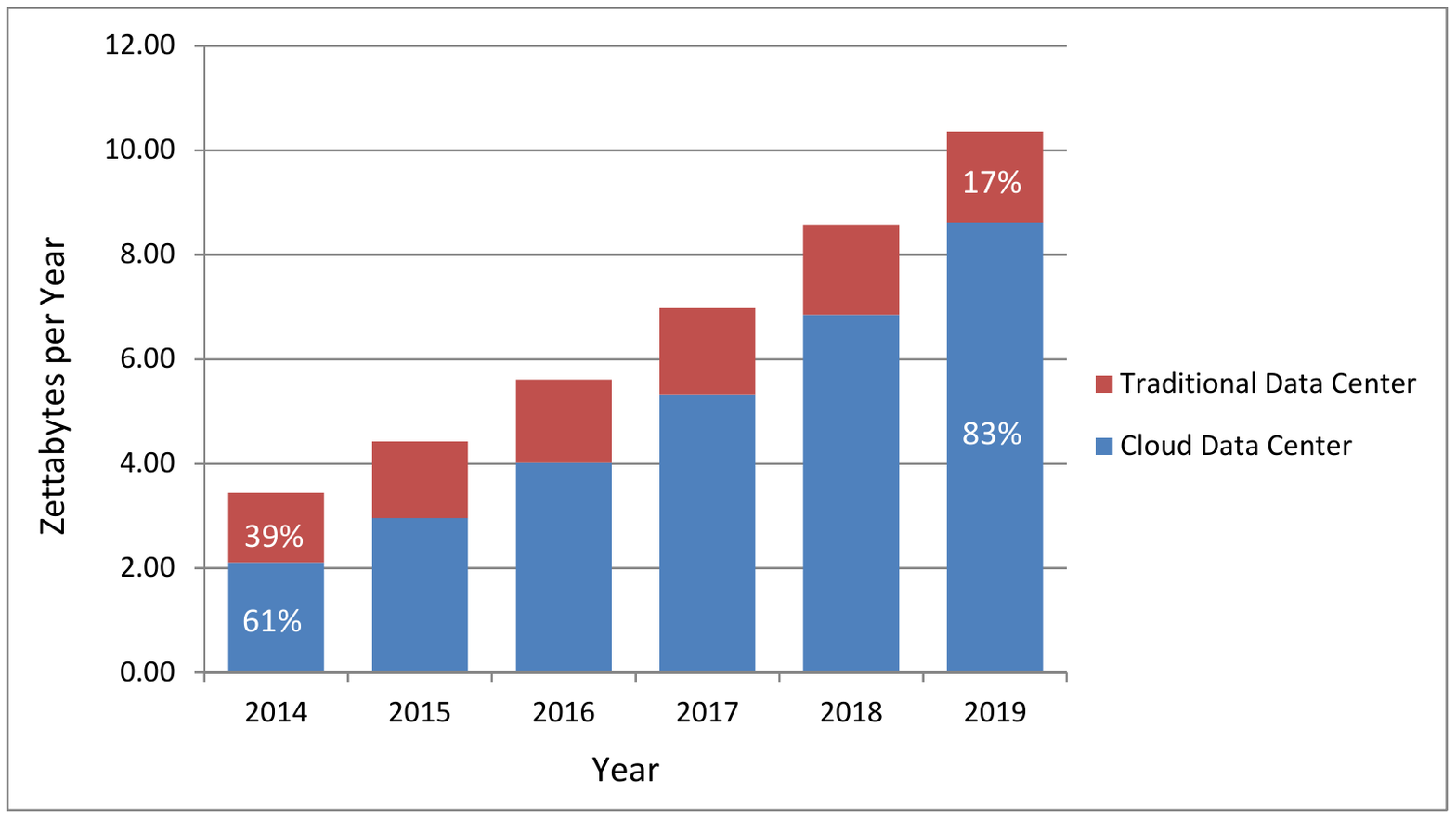}
\caption{Worldwide data center traffic growth (data source: Cisco).}
\label{chap1-fig-worldwide-data-center-traffic-growth}
\end{figure}

With the pragmatic realization of computing as a utility, Cloud Computing has recently emerged as a highly successful alternative information technology paradigm through the unique features of on-demand resource provisioning, pay-as-you-go business model, virtually unlimited amount of computing resources, and high reliability \citep{Buyya2009}. In order to meet the rapidly increasing demand for computing, communication, and storage resources, Cloud providers are deploying large-scale data centers comprising thousands of servers across the planet. These data centers are experiencing sharp rise in network traffic and a major portion of this traffic is constituted of the data communication within the data center. Recent report from Cisco Systems Inc. \citep{Cisco2015} demonstrates that the Cloud data centers will dominate the global data center traffic flow for the foreseeable future and its importance is highlighted by one of the top-line projections from this forecast that, by 2019, more than four-fifths of the total data center traffic will be Cloud traffic (Figure \ref{chap1-fig-worldwide-data-center-traffic-growth}). One important trait pointed out by the report is that a majority of the global data center traffic is generated due to the data communication within the data centers: in 2014, it was 75.4\% and it will be around 73.1\% in 2019.

This huge amount of intra-data center traffic is primarily generated by the application components that are correlated to each other, for example, the computing components of a composite application (e.g., MapReduce) writing data to the storage array after it has processed the data. This large growth of data center traffic may pose serious scalability problems for wide adoption of Cloud Computing. Moreover, by the way of continuously rising popularity of social networking sites, e-commerce, and Internet-based gaming applications, large amount of data processing has become an integral part of Cloud applications. Furthermore, scientific processing, multimedia rendering, workflow, and other massive parallel processing and business applications are being migrated to the Clouds due to the unique advantages of high scalability, reliability, and pay-per-use business model. Over and above, recent trend in Big Data computing using Cloud resources \citep{Assunccao2015} is emerging as a rapidly growing factor contributing to the rise of network traffic in Cloud data centers.

One of the key technological elements that have paved the way for the extreme success of Cloud Computing is virtualization. Modern data centers leverage various virtualization technologies (e.g., machine, network, and storage virtualization) to provide users an abstraction layer that delivers a uniform and seamless computing platform by hiding the underlying hardware heterogeneity, geographic boundaries, and internal management complexities \citep{Zhang2010}. By the use of virtualization, physical server resources are abstracted and shared through partial or full machine simulation by time-sharing, and hardware and software partitioning into multiple execution environments, known as \textit{Virtual Machines} (VMs), each of which runs as a complete and isolated system. It allows dynamic sharing and reconfiguration of physical resources in Cloud infrastructures that make it possible to run multiple applications in separate VMs having different performance metrics. It also facilitates Cloud providers to improve utilization of physical servers through VM multiplexing \citep{Meng2010a} and multi-tenancy, i.e., simultaneous sharing of physical resources of the same server by multiple Cloud customers. Furthermore, it enables on-demand resource pooling through which computing (e.g., CPU and memory), network, and storage resources are provisioned to customers only when needed \citep{Kusic2009}. By utilizing these flexible features of virtualization for provisioning physical resources, the scalability of data center network can be improved through minimization of network load imposed due to the deployment of customer applications.

\begin{figure}[!t]
\centering
\includegraphics[scale=0.5, trim=0cm 4cm 0cm 2cm]{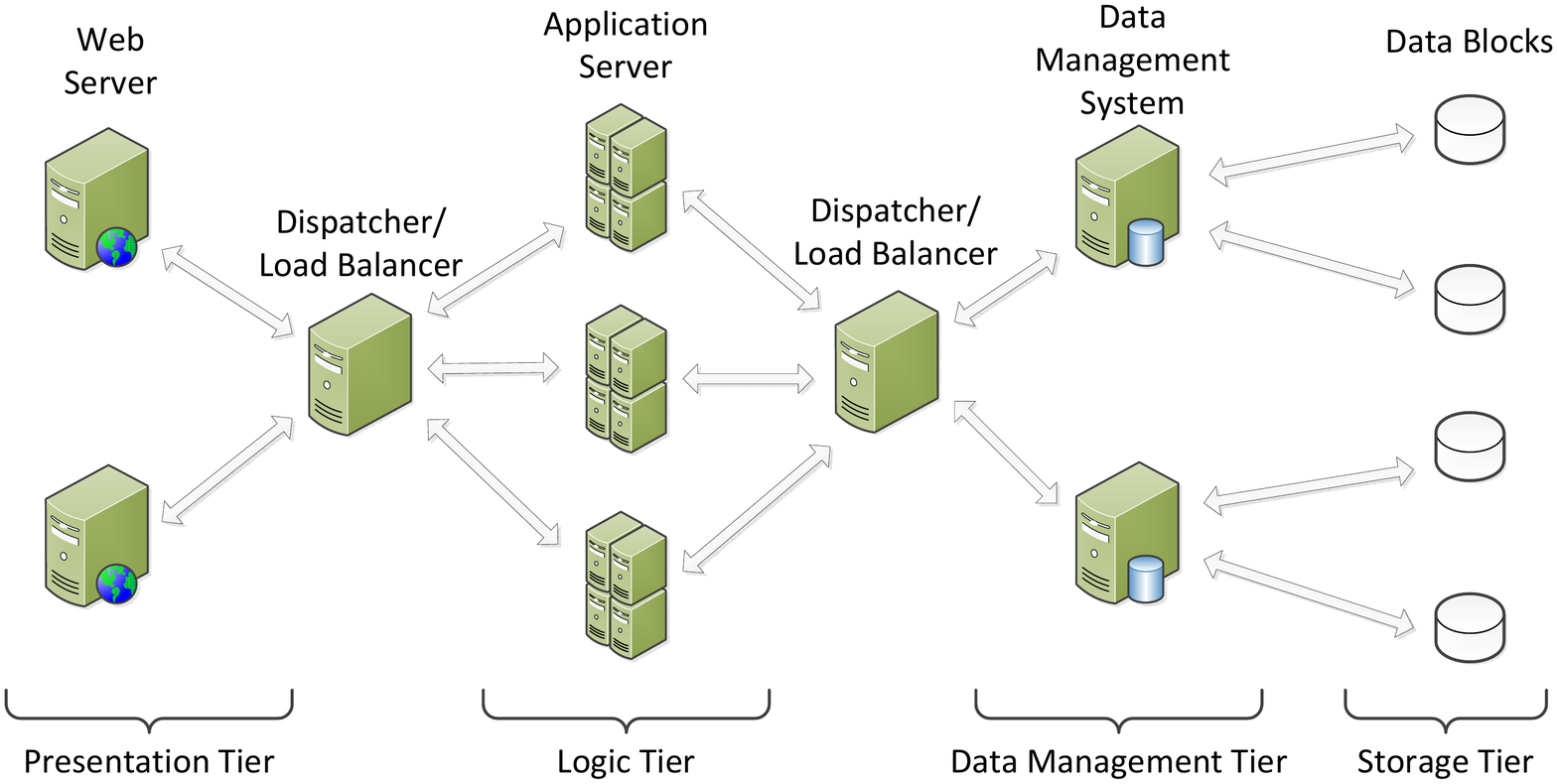}
\caption{Multi-tier application architecture.}
\label{chap4-fig-multi-tier-app-arc-1}
\end{figure}

On the other side, modern Cloud applications are dominated by multi-component applications such as multi-tier applications, massive parallel processing applications, scientific and business workflows, content delivery networks, and so on. These applications usually have multiple computing and associated data components. The computing components are usually delivered to customers in the form of VMs, such as Amazon EC2 Instances\footnote{Amazon EC2 - Virtual Server Hosting, 2016. https://aws.amazon.com/ec2/}, whereas the data components are delivered as data blocks, such as Amazon EBS\footnote{Amazon Elastic Block Store (EBS), 2016. https://aws.amazon.com/ebs/}. These computing components of such applications have specific service roles and are arranged in layers in the overall structural design of the application. For example, large enterprise applications are often modeled as 3-tier applications: the presentation tier (e.g., web server), the logic tier (e.g., application server), and the data tier (e.g., relational database) \citep{Urgaonkar2005}. The computing components (VMs) of such applications have specific communication requirements among themselves, as well as with the data blocks that are associated to those VMs (Figure \ref{chap4-fig-multi-tier-app-arc-1}). As a consequence, overall performance of such applications highly depends on the communication delays among the computing and data components. From the Cloud providers' perspective, optimization of network utilization of data center resources is tantamount to profit maximization. Moreover, efficient bandwidth allocation and reduction of data packet hopping through network devices (e.g., switches or routers) trim down the overall energy consumption of network infrastructure. On the other hand, Cloud consumers' concern is to receive guaranteed Quality of Service (QoS) of the delivered virtual resources, which can be assured through appropriate provisioning of requested resources.

Given the issues of sharp rise in network traffic in data centers, this paper addresses the scalability concern of data center network through a traffic-aware placement strategy of multi-component, composite application (in particular, VMs and data blocks) in virtualized data center that aims at optimizing the network traffic load incurred due to placement decision. Such placement decisions can be made during the application deployment phase in the data center. VM placement decisions focusing on other goals rather than network efficiency, such as energy consumption reduction (\citep{Feller2011}, \citep{Beloglazov2012}) and server resource utilization (\citep{Gao2013}, \citep{Ferdaus2014a}), often result in placements where VMs with high mutual traffic are placed in host servers with high mutual network cost. For example, one of our previous works \citep{Ferdaus2014a} on the placement of a cluster of VMs strives to consolidate the VMs into a minimal number of servers in order to reduce server resource wastage. By this process, unused servers can be kept into lower power states (e.g., suspended) so as to improve power efficiency of the data center. Since this approach does not consider inter-VM network communication patterns, such placement decisions can eventually result in locating VMs with high mutual network traffic in long distant servers, such as servers locating across the network edges. Several other VM placement works focusing on non-network objectives can be found in \citep{Wu2015}, \citep{Farahnakian2015}, \citep{Nguyen2014}, \citep{Corradi2014}, and \citep{Alboaneen2014}. With a network-focused analysis, it can be concluded that research works such as the above ones considered single-tier applications and VM clusters without consideration of mutual network communication within the application components or VMs. On the contrary, this paper focuses on placing mutually communicating components of applications (such as VMs and data blocks) in data center components (such as physical servers and storage devices) with lesser network cost so that network overhead imposed due to the application placement is minimized. With this placement goal, the best placement for two communicating VMs would be in the same server where they can communicate through memory copy, rather than using the physical network links. This paper effectively addresses network-focused placement problem of multi-tiered applications with components having mutual network communication rather than single-tiered ones. The significance of the network-focused placement of multi-tiered applications is evident from the experimental results presented later in Section \ref{chap4-sec-perf-eval}, where it is observed that an efficient non-network greedy placement algorithm, namely First Fit Decreasing (FFD), incurs higher network costs compared to the proposed network-aware placement heuristic.

Moreover, advanced hardware devices with combined capabilities are opening new opportunities for efficient resource allocation focusing on application needs. For example, Dell PowerEdge C8000 moduler servers are equipped with CPU, GPU, and storage components that can work as multi-function devices. Combined placement of application components with high mutual traffic (e.g., VMs and their associated data components) in such multi-function servers will effectively reduce the data transfer delay since the data accessed by the VMs reside in the same devices. Similar trends are found in high-end network switches (e.g., Cisco MDS 9200 Multiservice Switches) that come with additional built-in processing and storage capabilities. Reflecting on these technological development and multi-purpose devices, this paper has considered a generic approach in modeling computing, network, and storage elements in a data center so that placement algorithms can make efficient decision for application components placement in order to achieve the ultimate goal of network cost reduction.

This research work investigates the allocation, specifically on-demand placement of composite application components (modeled as an \textit{Application Environment}) requested by the customers to be deployed in Cloud data center focusing on network utilization, with consideration of computing, network, and storage resources capacity constraints of the data center. In particular, this paper has the following contributions:
\vspace{-2mm}
\begin{enumerate}
\item The Network-aware Application environment Placement Problem (NAPP) is formally defined as a combinatorial optimization problem with the objective of network cost minimization due to the placement. The proposed data center and application environment models are generic and are not restricted to any specific data center topology and application type or structure, respectively. 
\item Given the resource requirements and structure of the application environment to be deployed, and information on the current resource state of the data center, a Network- and Data location-aware Application environment Placement (NDAP) scheme is proposed. NDAP is a greedy heuristic that generates mappings for simultaneous placement of the computing and data components of the application into the computing and storage nodes of the data center, respectively, focusing on minimization of incurred network traffic, while respecting the computing, network, and storage capacity constraints of data center resources. While making placement decisions, NDAP strives to reduce the distance that data packets need to travel in the data center network, which in turn, helps to localize network traffic and reduces communication overhead in the upper layer network switches. 
\item Finally, performance evaluation of the proposed approach is conducted through elaborate simulation-based experimentation across multiple performance metrics and several scaling factors. The results suggest that the NDAP algorithm successfully improves network resource utilization through efficient placement of application components and outperforms compared algorithms significantly across all performance metrics.
\end{enumerate}

The proposed NDAP greedy heuristic for placement of application environments, while optimizing the overall network overhead, is addressing an important sub-problem of a much bigger multi-objective placement problem that simultaneously optimizes computing, storage, and communications resources. While many multi-objective works are available in the literature aiming at consolidation of the first two kinds of resources (computing and storage), works addressing all three kinds of resources are few and these works only considered placement of VMs in isolation. Development of NDAP is the first step in addressing the comprehensive optimization problem considering the placement of a group of closely-linked VMs, hereby termed as an application environment.

The remainder of this paper is organized as follows. A brief background on the related works is presented in Section \ref{chap4-sec-related-works}. Section \ref{chap4-sec-problem-stmt} formally defines the addressed application placement problem (NAPP) as an optimization problem, along with the associated mathematical models. The proposed network-aware, application placement approach (NDAP) and its associated algorithms are elaborately explicated in Section \ref{chap4-sec-proposed-solution}. Section \ref{chap4-sec-perf-eval} details the experiments performed and shows the results, together with their analysis. Finally, Section \ref{chap4-sec-con-future-work} concludes the paper with a summary of the contribution and future research directions.

\section{Related Work}
\label{chap4-sec-related-works}
During the past several years, a good amount of research works have been carried out in the area of VM scheduling, placement, and migration strategies in virtualized data centers, and more recently, focusing on Cloud data centers. A major portion of these works focus on servers resource utilization (\citep{Nguyen2014}, \citep{Gao2013}), energy-efficiency (\citep{Farahnakian2014}, \citep{Beloglazov2013}), and application performance (\citep{Gupta2013}, \citep{Calcavecchia2012}), and so on \citep{Ferdaus2014} in the context of large infrastructures. Recently, a handful of works are published in the area of VM placement and migration with focus on network resources that are briefly described below.

\citet{Kakadia2013} presented a VM grouping mechanism based on network traffic history within a data center at run-time and proposed a fast, greedy VM consolidation algorithm in order to improve hosted applications performance and optimize the network usage by saving internal bandwidth. Through simulation-based evaluation, the authors have shown that the proposed VM consolidation algorithm achieves better performance compared to traditional VM placement approaches, within an order to magnitude faster and requires much less VM migrations. \citet{Dias2012} addressed the problem of traffic concentration in data center networks by reallocating VMs in physical servers based on current traffic matrix and server resource usage. The authors proposed a scheme for partitioning server based on connectivity capacity and available computing resources, as well as VM clustering mechanism depending on the amount of data exchanged among the VMs. The proposed VM placement algorithm tries to find mappings for matching all the VM clusters in the server partitions, respecting the server resource capacity constraints. \citet{Shrivastava2011} proposed a topology-aware VM migration scheme for managing overloaded VMs considering the complete application context running on the VMs and the server resource capacity constraints. The goal of the proposed VM migration algorithm is to relocate overloaded VMs to physical servers so that the run-time network load within data center is minimized. Similar VM placement and relocation works can be found in \citep{Zhang2016}, \citep{Biran2012}, and \citep{Meng2010}, demand-based VM provisioning works for multi-component Cloud applications are presented in \citep{Srirama2014} and in \citep{Sahu2014}. All the above mentioned traffic-aware VM placement and consolidation works aim at run-time scenarios for relocating running VMs within the data center through VM migration. 


Several other recent VM placement and consolidation works have been proposed focusing on simultaneous optimization of energy and traffic load in data centers. \citet{Vu2014} addressed the issues of VM migration from underloaded and overloaded PMs at run-time and presented an offline algorithm for individual VM migration with the focus on traffic- and energy consumption reduction. Energy efficiency is achieved by consolidating VMs in high capacity servers as much as possible and traffic efficiency is achieved by migrating VMs near to communicating peer VMs. \citet{Wang2014} addressed the problem of unbalanced resource utilization and network traffic in data center during run-time, and proposed an energy-efficient and QoS-aware VM placement mechanism that groups the running VMs into partitions to reduce traffic communication across the data center, determines to server for migrating the VMs, and finally, uses the OpenFlow controller to assign paths to balance the traffic load and avoid congestion. \citet{Takouna2013} presented mechanisms for dynamically determining the bandwidth demand and communication patter of HPC and parallel applications in data center and reallocating the communicative VMs through live migration. The objective of the proposed approach is to reduce network link utilization and energy saving through aggregating communicative VMs. The authors have shown substantial improvement in data center traffic volume through simulation-based evaluation. \citet{Gao2016} addressed the problem of energy cost reduction under both server and network resource constraints within data center and proposed a VM placement strategy based on Ant Colony Optimization incorporating network resource factor with server resources. 

\citet{Huang2013} addressed the server overload problem and presented a three-stage joint optimization framework that minimizes the number of used servers in the data center in order to reduce power consumption, communication costs, and finally, a combined approach that focus on both the above goals through the use of VM migrations. \citet{Lloyd2014} investigated the problem of virtual resource provisioning and placement of service oriented applications through dynamic scaling in Cloud infrastructures and presented a server load-aware VM placement scheme that improves application performance and reduces resource cost. Similar multi-objective VM placement and migration works can also be found in \citep{Zhang2012}, \citep{Huang2012}, \citep{Wang2013}, and \citep{Song2012} that target optimization of energy consumption reduction, server resource utilization, and network usage. Given the fact that VM live migrations are costly operations \citep{Liu2013}, the above mentioned VM relocation strategies overlook the impact of necessary VM migrations and reconfiguration on hosted applications, physical servers and network devices. Further recent works on network-aware VM placement and migration can be found in \citep{Li2015}, \citep{Alharbi2016}, \citep{Cui2017}, \citep{Zhao2015}, and \citep{Wang2016}. A detailed taxonomy and survey on various existing network-aware VM management strategies can be found in our previous work \citep{Ferdaus2015}.

Contrary to the above mentioned works, this paper addresses the problem of network efficient, on-demand placement of composite applications consisting of multiple VMs and associated data components, along with inter-component communication pattern, in a data center consisting of both computing servers and storage devices. The addressed problem does not involve VM migrations since the placement decision is taken during the application deployment phase. Recently, \citet{Georgiou2013} have addressed the benefit of user-provided hints on inter-VM communication during the online VM cluster placement and proposed two placement heuristics utilizing the properties of PortLand network topology \citep{Mysore2009}. However, this work does not involve any data component for VM-cluster specification. On the other hand, both of the proposed composite, multi-tier application and data center models of this paper are generic and are not restricted to any particular application or data center topology. Data location-aware VM placement works can be found in \citep{Piao2010} and in \citep{Korupolu2009}; however, these works modeled the applications as single instance of VM, which is an oversimplified view of today's Cloud or Internet applications that are mostly composed of multiple computing and storage entities in multi-tier structure with strong communication correlations among the components. In order to reflect on this, this paper investigates a much wider VM communication model by considering placement of Application Environments, each involving a number of VMs and associated data blocks with sparse communication links between them.

\section{Problem Statement}
\label{chap4-sec-problem-stmt}
\begin{figure}[t]
\centering
\includegraphics[scale=0.5, trim=1cm .5cm 0cm 1cm]{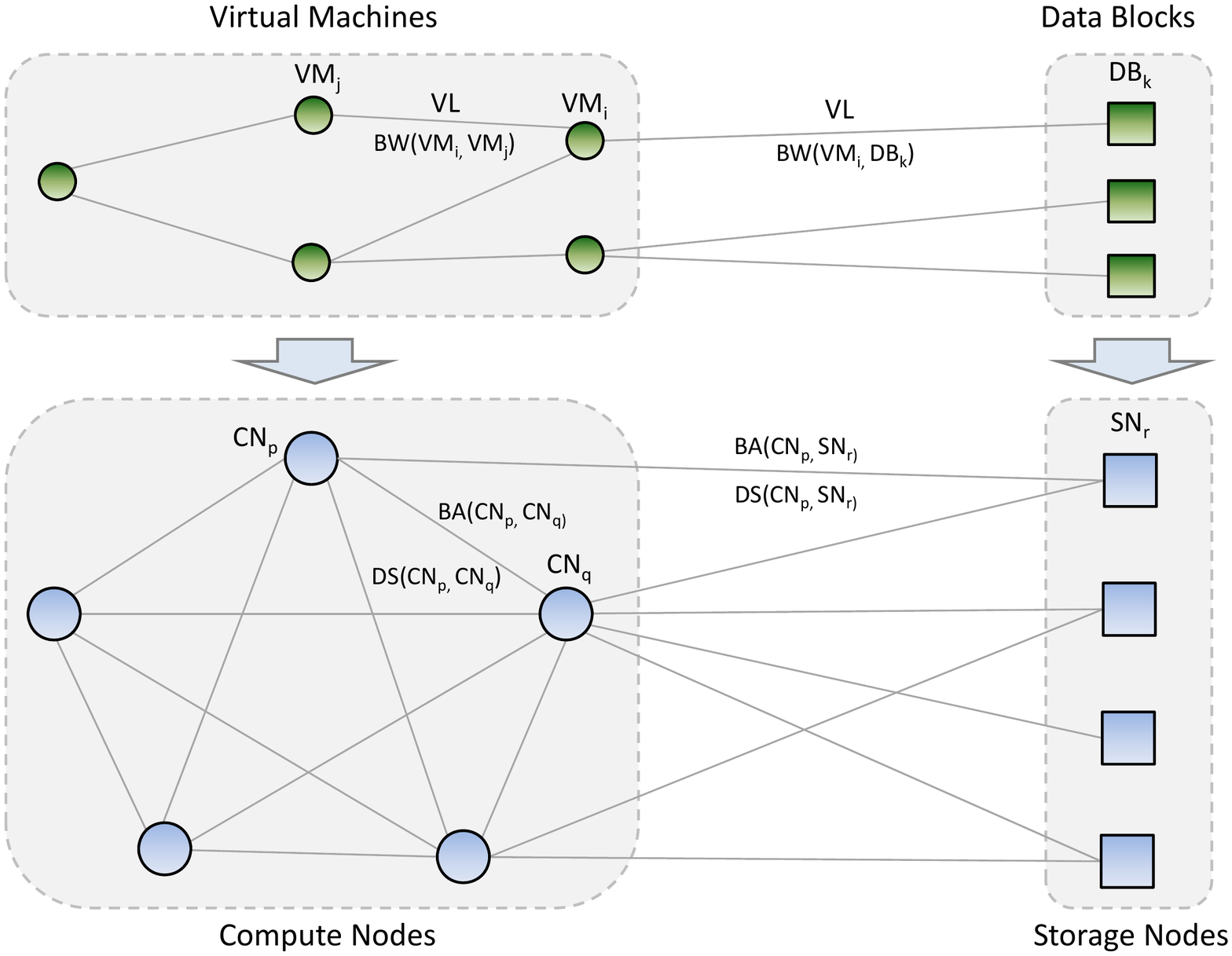}
\caption{Application environment placement on data center.}
\label{chap4-fig-application-placement-problem-1}
\end{figure}

While deploying composite applications in Cloud data centers, such as multi-tier or workflow applications, customers request multiple computing VMs in the form of a VM cluster or a Virtual Private Cloud and multiple Data Blocks (DBs). These computing VMs have specific traffic flow requirements among themselves, as well as with the data blocks. The remainder of this section formally defines such composite application environment placement as an optimization problem. Figure \ref{chap4-fig-application-placement-problem-1} presents a visual representation of the application placement in data center and Table \ref{chap4-tab-notations-1} provides the various notations used in the problem definition and proposed solution. 

\begin{table}[!t]
\centering
\caption{Notations and their meanings}
\label{chap4-tab-notations-1}
\begin{tabular}{|l|l|}
\hline
\multicolumn{1}{|c|}{\text{$Notation$}} & \multicolumn{1}{|c|}{\text{$Meaning$}} \\
\hline
$VM$	& Virtual Machine 	\\
$DB$	& Data Block	\\
$AN$	& AE node (either a VM or a DB)	\\
$VMS$	& Set of VMs in an AE	\\
$DBS$	& Set of DBs in an AE	\\
$ANS$	& Set of ANs ($ANS = \{VMS \cup DBS\}$)	in an AE	\\
$N_{v}$	& Total number of VMs in an AE 	\\
$N_{d}$	& Total number of DBs in an AE 	\\

$VL$	& Virtual Link	\\
$VCL$	& Virtual Computing Link that connects two VMs	\\
$VDL$	& Virtual Data Link	that connects a VM and a DB	\\
$vclList$	& Ordered list of VCLs in an AE 	\\
$vdlList$	& Ordered list of VDLs in an AE 	\\
$N_{vc}$	& Total number of VCLs in an AE		\\
$N_{vd}$	& Total number of VDLs in an AE		\\
$N_{vn}$	& Average number of NTPP VLs of a VM or a DB	\\
$BW(VM_{i}, VM_{j})$	& Bandwidth demand between $VM_{i}$ and $VM_{j}$	\\
$BW(VM_{i}, DB_{k})$	& Bandwidth demand between $VM_{i}$ and $DB_{k}$	\\
\hline
$CN$		& Computing Node	\\
$SN$		& Storage Node	\\
$DN(AN)$	& DC node where AN is placed	\\
$CNS$		& Set of CNs in a DC	\\
$SNS$		& Set of SNs in a DC	\\
$N_{c}$		& Total number of CNs in a DC 	\\
$N_{s}$		& Total number of SNs in a DC 	\\
$cnList$	& Ordered list of CNs in a DC 	\\
$snList$	& Ordered list of SNs in a DC 	\\

$PL$		& Physical network Link	\\
$PCL$		& Physical Computing Link that connects two CNs	\\
$PDL$		& Physical Data Link that connects a CN and a SN	\\

$DS(CN_{p}, CN_{q})$	& Network distance between $CN_{p}$ and $CN_{q}$	\\
$DS(CN_{p}, SN_{r})$	& Network distance between $CN_{p}$ and $SN_{r}$	\\
$BA(CN_{p}, CN_{q})$	& Available bandwidth between $CN_{p}$ and $CN_{q}$	\\
$BA(CN_{p}, SN_{r})$	& Available bandwidth between $CN_{p}$ and $SN_{r}$	\\

\hline
\end{tabular}
\end{table}

\subsection{Formal Definition}
\label{chap4-sec-formal-def}
An \emph{Application Environment} is defined as $AE=\lbrace VMS, DBS \rbrace$, where VMS is the set of requested VMs: $VMS=\lbrace VM_{i}: 1 \leq i \leq N_{v} \rbrace$ and DBS is the set of requested DBs: $DBS=\lbrace DB_{k}: 1 \leq k \leq N_{d} \rbrace$. Each VM $VM_{i}$ has specification of its CPU and memory demands represented by $VM_{i}^{cpu}$ and $VM_{i}^{mem}$, respectively, and each DB $DB_{k}$ has specification of its storage resource demand denoted by $DB_{k}^{str}$.

Data communication requirements between any two VMs, and between a VM and a DB are specified as \textit{Virtual Links} (VLs) between $\langle VM,VM \rangle$ pairs and $\langle VM,DB \rangle$ pairs, respectively, during AE specification and deployment. The bandwidth demand or traffic load between $VM_{i}$ and $VM_{j}$ is represented by $BW(VM_{i}, VM_{j})$. Similarly, the bandwidth demand between $VM_{i}$ and $DB_{k}$ is represented by $BW(VM_{i}, DB_{k})$. These bandwidth requirements are provided as user input along with the VM and DB specifications. 

A \emph{Data Center} is defined as $DC=\lbrace CNS, SNS \rbrace$ where CNS is the set of computing nodes (e.g., physical servers or computing components of a multi-function storage device) in DC: $CNS=\lbrace CN_{p}: 1 \leq p \leq N_{c} \rbrace$ and SNS is the set of storage nodes: $SNS=\lbrace SN_{r}: 1 \leq r \leq N_{s} \rbrace$. For each computing node $CN_{p}$, the available CPU and memory resource capacities are represented by $CN_{p}^{cpu}$ and $CN_{p}^{mem}$, respectively. Here available resources mean the remaining usable resources of a CN that may have already hosted other VMs that are consuming the rest of the resources. Similarly, for each storage node $SN_{r}$, the available storage resource capacity is represented by $SN_{r}^{str}$.

Computing nodes and storage nodes are interconnected through \textit{Physical Links} (PLs) in the data center communication network. PL distance and available bandwidth between two computing nodes $CN_{p}$ and $CN_{q}$ are denoted by $DS(CN_{p}, CN_{q})$ and $BA(CN_{p}, CN_{q})$, respectively. Similarly, PL distance and available bandwidth between a computing node $CN_{p}$ and a storage node $SN_{r}$ are represented by $DS(CN_{p}, SN_{r})$ and $BA(CN_{p}, SN_{r})$, respectively. PL distance can be any practical measure, such as link latency, number of hops or switches, and so on. Also, the proposed model does not restrict the data center to a fixed network topology. Thus, the network distance $DS$ and available bandwidth $BA$ models are generic and different model formulations focusing on any particular network topology or architecture can be readily applied in the optimization framework and proposed solution. In the experiments, the number of hops or switches between any two data center nodes is used as the only input parameter for $DS$ function in order to measure the PL distance. Although singular distances between $\langle CN,CN \rangle$ and $\langle CN,SN \rangle$ pairs are used in the experiments, network link redundancy and multiple communication paths in data center can be incorporated in the proposed model and placement algorithm by appropriately defining distance function ($DS$) and available bandwidth function ($BA$), respectively.

Furthermore, $DN(VM_{i})$ denotes the computing node where $VM_{i}$ is currently placed, otherwise if $VM_{i}$ is not already placed, $DN(VM_{i}) = null$. Similarly, $DN(DB_{k})$ denotes the storage node where $DB_{k}$ is currently placed.

The network cost of placing $VM_{i}$ in $CN_{p}$ and $VM_{j}$ in $CN_{q}$ is defined as the following:
\begin{equation}
\label{chap4-eq-cost-1}
	Cost(VM_{i}, CN_{p}, VM_{j}, CN_{q}) =  BW(VM_{i}, VM_{j}) \times DS(CN_{p}, CN_{q}).
\end{equation}

Likewise, the network cost of placing $VM_{i}$ in $CN_{p}$ and $DB_{k}$ in $SN_{r}$ is defined as the following:
\begin{equation}
\label{chap4-eq-cost-2}
	Cost(VM_{i}, CN_{p}, DB_{k}, SN_{r}) =  BW(VM_{i}, DB_{k}) \times DS(CN_{p}, SN_{r}).
\end{equation}

Given the AE to deploy in the DC, the objective of the NAPP problem is to find placements for VMs and DBs in CNs and SNs, respectively, in such a way that the overall network cost or communication overhead due to the AE deployment is minimized. Thus, the \emph{Objective Function f} is defined as the following:

\begin{align}
\label{chap4-eq-ndap-obj-func-1}
\begin{split}
	\minimize_{\substack{\forall i: DN(VM_{i}) \\ \forall k: DN(VM_{k})}}  f(AE, DC) = \sum_{i=1}^{N_{v}} \bigg( & \sum_{j=1}^{N_{v}} Cost(VM_{i}, DN(VM_{i}), VM_{j}, DN(VM_{j})) +  \\
								& \sum_{k=1}^{N_{d}} Cost(VM_{i}, DN(VM_{i}), DB_{k}, DN(DB_{k})) \bigg).
\end{split}
\end{align}

The above AE placement is subject to the constraints that the available resource capacities of any CN and SN are not violated:
\begin{align}
\label{chap4-eq-cn-cpu-constr-1}
\forall p: \sum_{\forall i: DN(VM_{i}) =	CN_{p}} VM_{i}^{cpu} & \leq CN_{p}^{cpu}. &
\end{align}

\begin{align}
\label{chap4-eq-cn-mem-constr-1}
\forall p: \sum_{\forall i: DN(VM_{i}) = CN_{p}} VM_{i}^{mem} & \leq CN_{p}^{mem}. &
\end{align}

\begin{align}
\label{chap4-eq-sn-str-constr-1}
\forall r: \sum_{\forall k: DN(DB_{k}) = SN_{r}} DB_{k}^{str} & \leq SN_{r}^{str}. &
\end{align}

Furthermore, the sum of the bandwidth demands of the VLs that are placed on each PL must be less or equal to the available bandwidth of the PL:

\begin{align}
\label{chap4-eq-cn-cn-ba-constr-1}
\begin{split}
	\forall p \forall q: {\;} BA(CN_{p}, CN_{q}) \geq {} \sum_{\forall i: DN(VM_{i}) = CN_{p}} {  }\sum_{\forall j: DN(VM_{j}) = CN_{q}} BW(VM_{i}, VM_{j}).
\end{split}
\end{align}

\begin{align}
\label{chap4-eq-cn-sn-ba-constr-1}
\begin{split}
	\forall p \forall r: {\;} BA(CN_{p}, SN_{r}) \geq {} \sum_{\forall i: DN(VM_{i}) = CN_{p}} {  } \sum_{\forall k: DN(DB_{k}) = SN_{r}} BW(VM_{i}, DB_{k}).
\end{split}
\end{align}

Given that every VM and DB placement fulfills the above mentioned constraints (Eq. \ref{chap4-eq-cn-cpu-constr-1}--\ref{chap4-eq-cn-sn-ba-constr-1}), the NAPP problem defined by objective function $f$ (Eq. \ref{chap4-eq-ndap-obj-func-1}) is explained as: among all possible feasible placements of VMs and DBs in AE, the placement that has minimum cost is the optimal solution. Thus, NAPP falls in the category of combinatorial optimization problem. In particular, it is an extended form of the \textit{Quadratic Assignment Problem} (QAP) \citep{Loiola2007}, which is proven to be computationally $\mathcal{NP-}$hard \citep{Burkard1998}.

\section{Proposed Solution}
\label{chap4-sec-proposed-solution}
The proposed network-aware VM and DB placement approach (NDAP) tries to place the VLs in such a way that network packets need to travel short distances. For better explanation of the solution approach, the above described models of AE and DC are extended by adding few other notations. 

Every AE node is represented by $AN$ which can either be a VM or a DB, and the set of all ANs in an AE is represented by $ANS$. Every $VL$ can be either a \emph{Virtual computing Link (VCL)}, i.e., $VL$ between two VMs or a \emph{Virtual Data Link (VDL)}, i.e., VL between a VM and a DB. The total number of VCL and VDL in an AE is represented by $N_{vc}$ and $N_{vd}$, respectively. All the VCLs and VDLs are maintained in two ordered lists $vclList$ and $vdlList$, respectively. While VM-VM communication (VCL) and VM-DB communication (VDL) may be considered closely related, they differ in terms of actor and size. As only a VM can initiate communications, VCL supports an "active" duplex link while VDL supports a "passive" duplex link. More distinctly, bandwidth demands of VDLs are multiple orders larger than the same of VCLs.

Every DC node is represented by $DN$ which can either be a $CN$ or a $SN$. All the CNs and SNs in a DC are maintained in two ordered lists $cnList$ and $snList$, respectively. Every $PL$ can be either a \emph{Physical computing Link (PCL)}, i.e., $PL$ between two CNs or a \emph{Physical Data Link (PDL)} i.e., $PL$ between a CN and a SN. 

The proposed NDAP algorithm is a greedy heuristic that first sorts the $vdlList$ and $vclList$ in decreasing order of the bandwidth demand of VDLs and VCLs. Then, it tries to places all the VDLs from $vdlList$, along with any associated VCLs to fulfill placement dependency, on the feasible PDLs and PCLs, and their associated VMs and DBs in CNs and SNs, respectively, focusing on the goal of minimizing the incurred network cost due to placement of all the VDLs and associated VCLs. Finally, NDAP tries to place the remaining VCLs from $vclList$ on PCLs, along with their associated VMs and DBs in CNs and SNs, respectively, again targeting on reducing the incurred network cost.

As mentioned in Section \ref{chap4-sec-problem-stmt}, NAPP is in fact an $\mathcal{NP-}$hard combinatorial optimization problem similar to QAP and \citet{Sahni1976} have shown that even finding an approximate solution for QAP within some constant factor from the optimal solution cannot be done in polynomial time unless $\mathcal{P=NP}$. Considering the fact that greedy heuristics are relatively fast, easy to understand and implement, and very often used as an effective solution approach for $\mathcal{NP-}complete$ problems, this paper proposes NDAP greedy heuristic as a solution for the NAPP problem.

A straight forward placement of an individual VL (either VDL or VCL) on a preferred PL is not always possible since one or both of its ANs can have \textit{Peer ANs} connected by \textit{Peer VLs} (Figure \ref{chap4-fig-possible-VL-placement-scenarios-1}(a)). At any point during an AE placement process, a VL can have Peer ANs that are already placed. The peer VLs that have already-placed peer ANs is termed as \textit{need-to-place peer VLs} (NTPP VLs), indicating the condition that placement of any VL also needs to perform simultaneous placement of its NTPP VLs, and the average number of NTPP VLs for any VM or DB is denoted by $N_{vn}$. The maximum value of $N_{vn}$ can be $N_{v} + N_{d} - 1$ which indicates that the corresponding VM or DB has VLs with all the other VMs and DBs in the AE. Since, for any VL placement, the corresponding placement of its NTPP VLs is an integrated part of the NDAP placement strategy, firstly the VL placement feasibility part of the NDAP algorithm is presented in the following subsection. Afterwards, the next four subsections describe other constituent components of the NDAP algorithm. Finally, a detailed description of the final NDAP algorithm is provided along with the pseudocode.

\subsection{VL Placement Feasibility}
\label{chap4-sec-vl-place-feas}
During the course of AE placement, when NDAP tries to place a $VL$ that has one or both of its ANs not placed yet (i.e., $DN(AN) = null$), then a feasible placement for the $VL$ needs to ensure that (1) the $VL$ itself is placed on a feasible PL, (2) its ANs are placed on feasible DNs, and (3) all the NTPP VLs are placed on feasible PLs. 

\begin{figure}[t]
\centering
\includegraphics[scale=0.6, trim=1.2cm .5cm 0cm 0.3cm]{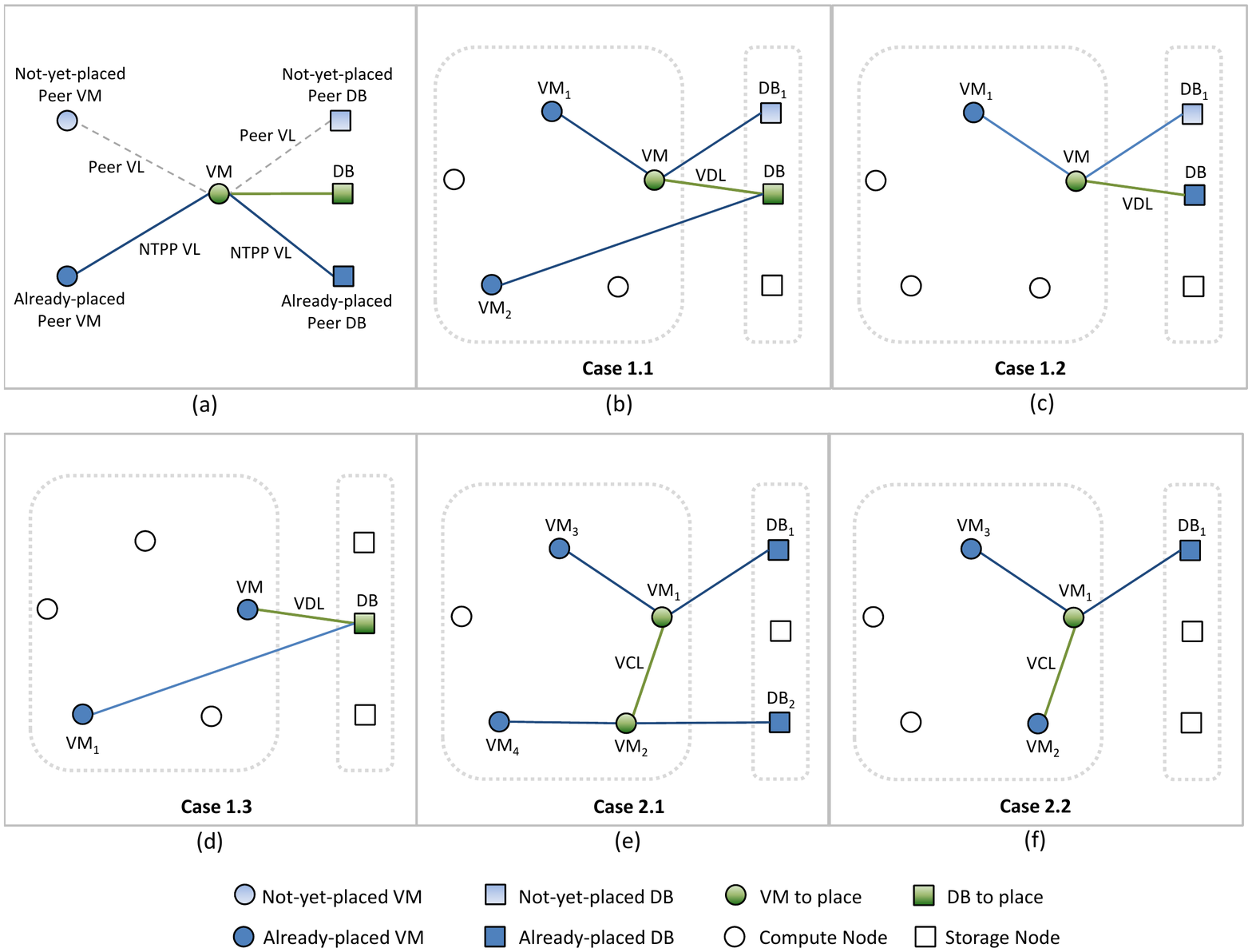}
\caption{(a) Peer VL and NTPP VL, and (b-f) Five possible VL placement scenarios.}
\label{chap4-fig-possible-VL-placement-scenarios-1}
\end{figure}

Depending on the type of $VL$ and the current placement status of its ANs, five different cases may arise that are presented below. The NDAP placement algorithm handles these five cases separately. Figure \ref{chap4-fig-possible-VL-placement-scenarios-1}(b)-(f) provide a visual representation of the five cases where the $VL$ to place is shown as solid green line and its NTPP VLs are shown as solid blue lines.

\textbf{VDL Placement: } When trying to place a $VDL$, any of the following three cases may arise:

\textit{Case 1.1:} Both the $VM$ and $DB$ are not placed yet and their peers $VM_{1}$, $DB_{1}$, and $VM_{2}$ are already placed (Figure \ref{chap4-fig-possible-VL-placement-scenarios-1}(b)).

\textit{Case 1.2:} $DB$ is placed but $VM$ is not placed yet and $VM$'s peers $VM_{1}$ and $DB_{1}$ are already placed (Figure \ref{chap4-fig-possible-VL-placement-scenarios-1}(c)).

\textit{Case 1.3:} $VM$ is placed but $DB$ is not placed yet and $DB$'s peer $VM_{1}$ is already placed (Figure \ref{chap4-fig-possible-VL-placement-scenarios-1}(d)). 

\textbf{VCL Placement: } In case of $VCL$ placement, any of the following two cases may arise:

\textit{Case 2.1:} Both the VMs ($VM_{1}$ and $VM_{2}$) are not placed yet and their peers $VM_{3}$, $DB_{1}$, $VM_{4}$, and $DB_{2}$ are already placed (Figure \ref{chap4-fig-possible-VL-placement-scenarios-1}(e)). 

\textit{Case 2.2:} Only one of the VMs is already placed and its peers $VM_{3}$ and $DB_{1}$ are already placed (Figure \ref{chap4-fig-possible-VL-placement-scenarios-1}(f)).

In all the above cases, placement feasibility of the NTPP VDLs and VCLs of the not-yet-placed VMs and DBs must be checked against the corresponding PDLs and PCLs, respectively (Eq. \ref{chap4-eq-cn-cn-ba-constr-1} \& \ref{chap4-eq-cn-sn-ba-constr-1}). 

\subsection{Feasibility and Network Cost of VM and Peer VLs Placement}
\label{chap4-sec-feas-netcost-vm-peer}
When NDAP tries to place a VM in a CN, it is feasible when (1) the computing and memory resource demands of the VM can be fulfilled by the remaining computing and memory resource capacities of the CN, and (2) the bandwidth demands of all the NTPP VLs can be satisfied by the available bandwidth capacities of the corresponding underlying PLs (Figure \ref{chap4-fig-vdl-vcl-placement-1}(a)):
\begin{equation}
\label{chap4-eq-vm-place-feas-1}
VMPeerFeas(VM, CN)=
   \begin{cases}
   1, & \text{if Eq. \ref{chap4-eq-cn-cpu-constr-1} \& \ref{chap4-eq-cn-mem-constr-1} holds and, } DN(AN) \neq \text{null and } \\ 
      & BW(VM,AN) \leq BA(CN, DN(AN)) { \text{ for } \forall AN}; \\
   0, & \text{otherwise}.
  \end{cases}
\end{equation}

When NDAP tries to place two VMs ($VM_{1}$ and $VM_{2}$) in a single CN, it is feasible when (1) the combined computing and memory resource demands of the two VMs can be fulfilled by the remaining computing and memory resource capacities of the CN, and (2) the bandwidth demands of all the NTPP VLs of both the VMs can be satisfied by the available bandwidth capacities of the corresponding underlying PLs:
\begin{equation}
\label{chap4-eq-vm-place-feas-2}
VMPeerFeas(VM_{1}, VM_{2}, CN)=
   \begin{cases}
   1, & \text{if Eq. \ref{chap4-eq-cn-cpu-constr-1} \& \ref{chap4-eq-cn-mem-constr-1} holds for } (VM_{1}+VM_{2}) \text{ and, }  \\
   	  & \forall AN: DN(AN) \neq \text{null and, } \\
   	  & BW(VM_{1},AN) + BW(VM_{2},AN) \leq BA(CN, DN(AN)); \\
   0, & \text{otherwise}.
  \end{cases}
\end{equation}

The network cost of a VM placement is measured as the accumulated cost of placing all of its NTPP VLs:
\begin{equation}
\label{chap4-eq-vm-place-cost-1}
VMPeerCost(VM,CN) = \sum_{\forall AN: \,DN(AN) \neq null \wedge BW(VM, AN) > 0} Cost(VM, CN, AN, DN(AN)).
\end{equation}

\subsection{Feasibility and Network Cost of DB and Peer VLs Placement}
\label{chap4-sec-feas-netcost-db-peer}
\begin{figure}[t]
\centering
\includegraphics[scale=0.55, trim=2.5cm 0cm 0cm 0cm]{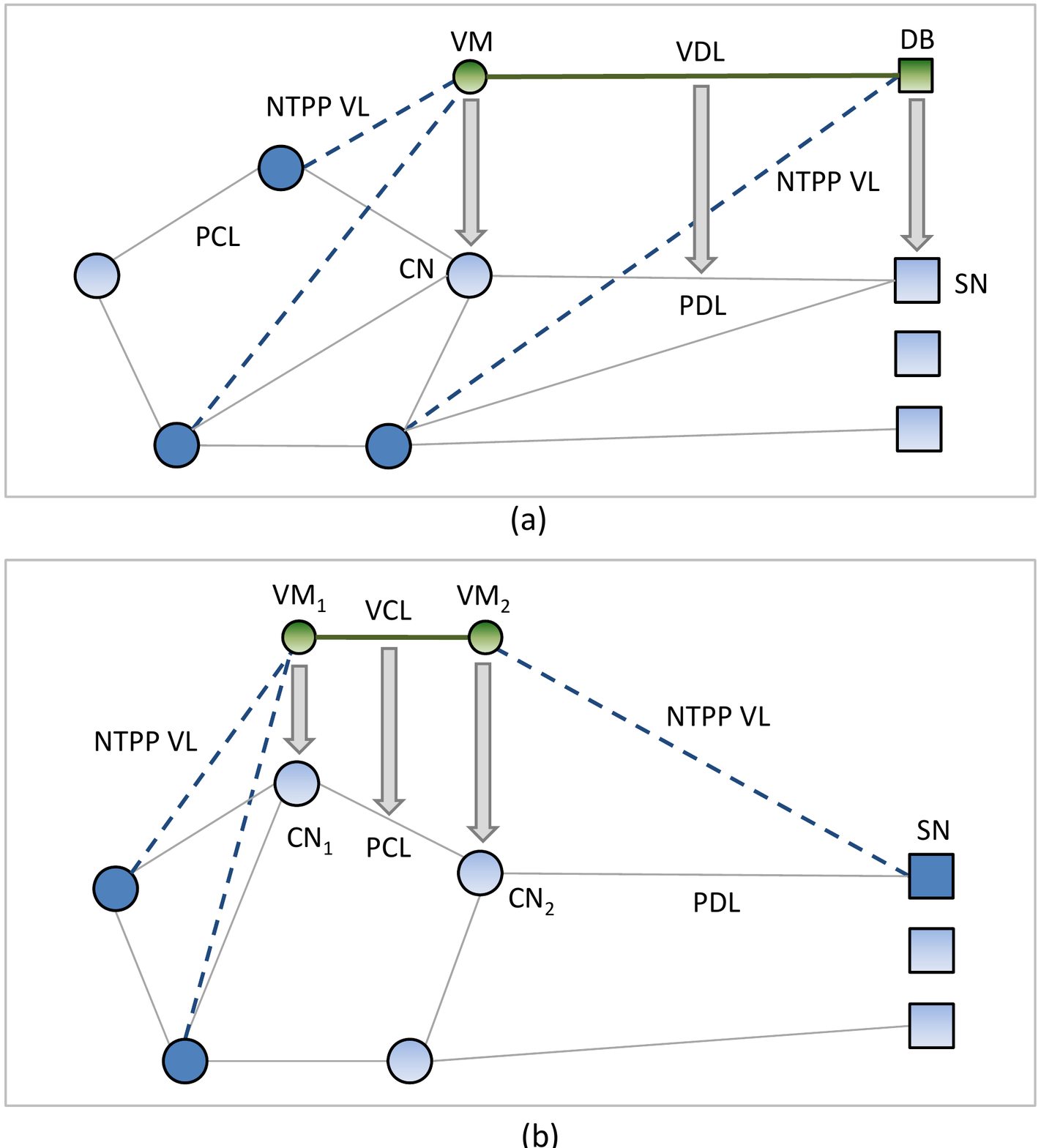}
\caption{Placement of (a) VDL and (b) VCL along with NTPP VLs.}
\label{chap4-fig-vdl-vcl-placement-1}
\end{figure}

When trying to place a DB in a SN, it is feasible when (1) the storage resource demand of the DB can be fulfilled by the remaining storage resource capacity of the SN, and (2) the bandwidth demands of the NTPP VLs can be satisfied by the available bandwidth capacities of corresponding underlying PLs (Figure \ref{chap4-fig-vdl-vcl-placement-1}(a)):
\begin{equation}
\label{chap4-eq-db-place-feas-1}
DBPeerFeas(DB, SN)=
   \begin{cases}
   1, & \text{if Eq. \ref{chap4-eq-sn-str-constr-1} holds and, } DN(AN) \neq \text{null and } \\ 
      & BW(AN,DB) \leq BA(DN(AN),SN) { \text{ for } \forall AN}; \\
   0, & \text{otherwise}.
  \end{cases}
\end{equation}

The network cost of any DB placement is measured as the total cost of placing all of its NTPP VLs:
\begin{equation}
\label{chap4-eq-db-place-cost-1}
DBPeerCost(DB,SN) = \sum_{\forall AN: \,DN(AN) \neq null \wedge BW(AN, DB) > 0} Cost(AN, DN(AN), DB, SN).
\end{equation}

\subsection{VM and Peer VLs Placement}
\label{chap4-sec-vm-peer-place}
Algorithm \ref{chap4-alg-placeVMandpeerVLs-1} shows the subroutine for placing a $VM$ and its associated NTPP VLs. Firstly, the $VM$-to-$CN$ placement is accomplished by reducing the available CPU and memory resource capacities of the $CN$ by the amount of CPU and memory resource requirements of the $VM$ and setting the $CN$ as the DC node of the $VM$ [line 1]. Then, for each already-placed peer $AN$ of $VM$ (i.e., any $AN$ that has non-zero traffic load with $VM$ and $DN(AN) \neq null$), it is checked if the selected $CN$ is different from the computing node where the peer $AN$ is placed, in which case the available bandwidth capacity of the PL that connects the selected $CN$ and $DN(AN)$ is reduced by the amount of the bandwidth demand of the corresponding NTPP VL [lines 2--4]. In those cases where the selected $CN$ is the computing node where the peer $AN$ is placed, the $VM$ can communicate with the peer $AN$ through memory copy instead of passing packet through physical network links. Afterwards, the NTPP $VL$ is removed from the $vclList$ or $vdlList$, depending on whether it is a VCL or VDL, respectively, in order to indicate that it is now placed [lines 5--7].
\begin{algorithm}[t]
\textbf{Input:} $VM$ to place, $CN$ where $VM$ is being placed, set of all ANs $ANS$, $vclList$, and $vdlList$. \\
\textbf{Output:} $VM$-to-$CN$ and $VL$-to-$PL$ placements.
\begin{algorithmic}[1]
\STATE $CN^{cpu} \leftarrow CN^{cpu} - VM^{cpu}; CN^{mem} \leftarrow CN^{mem} - VM^{mem}; DN(VM) \leftarrow CN;$
\FOR{each $AN \in ANS$}
	\IF {$BW(VM, AN) > 0 \wedge DN(AN) \neq null$}	
		\STATE \textbf{if} $DN(AN) \neq CN$ \textbf{then} $BA(CN, DN(AN)) \leftarrow BA(CN, DN(AN)) - BW(VM, AN);$ \textbf{endif}
		\STATE $VL	\leftarrow virtualLink(VM, AN);$
		\STATE \textbf{if} $VL$ is a $VCL$ \textbf{then} $vclList.remove(VL);$ \
		\STATE \textbf{else} $vdlList.remove(VL);$ 
		\STATE \textbf{endif}
	\ENDIF
\ENDFOR
\end{algorithmic}
\caption{PlaceVMandPeerVLs}
\label{chap4-alg-placeVMandpeerVLs-1}
\end{algorithm}

\subsection{DB and Peer VLs Placement}
\label{chap4-sec-db-peer-place}
Algorithm \ref{chap4-alg-placeDBandpeerVLs-1} shows the subroutine for placing a $DB$ in a $SN$ and its associated NTPP VLs. Firstly, the $DB$-to-$SN$ placement is performed by reducing the available storage capacity of the $SN$ by the amount of the storage requirements of the $DB$ and by setting the $SN$ as the DC node of $DB$ [line 1]. Then, for every already-placed peer $AN$ of $DB$ (i.e., any $AN$ that has non-zero traffic load with $DB$ and $DN(AN) \neq null$), the available bandwidth capacity of the PDL that connects the selected $SN$ and $DN(AN)$ is reduced by the amount of the NTPP $VL$'s bandwidth requirement and the NTPP $VL$ is removed from the $vdlList$ to mark that it is now placed [lines 2--6].
\begin{algorithm}[t]
\textbf{Input:} $DB$ to place, $SN$ where $DB$ is being placed, set of all ANs $ANS$, and $vdlList$. \\
\textbf{Output:} $DB$-to-$SN$ and $VL$-to-$PL$ placements.
\begin{algorithmic}[1]
\STATE $SN^{str} \leftarrow SN^{str} - DB^{str}; DN(DB) \leftarrow SN;$
\FOR{each $AN \in ANS$}
	\IF {$BW(AN, DB) > 0 \wedge DN(AN) \neq null$}	
		\STATE $BA(DN(AN), SN) \leftarrow BA(DN(AN), SN) - BW(AN, DB); VL \leftarrow virtualLink(AN, DB); vdlList.remove(VL);$
	\ENDIF
\ENDFOR
\end{algorithmic}
\caption{PlaceDBandPeerVLs}
\label{chap4-alg-placeDBandpeerVLs-1}
\end{algorithm}

\subsection{NDAP Algorithm}
\label{chap4-sec-ndap-alg}
The pseudocode of the final NDAP algorithm is presented in Algorithm \ref{chap4-alg-ndap-1}. It receives the $DC$ and $AE$ as input and returns the network cost incurred due to the $AE$ placement. NDAP begins by performing necessary initialization and sorting the $vdlList$ and $vclList$ in decreasing order of their VLs' bandwidth demands [line 1]. Afterwards, it iteratively takes the first VDL from $vdlList$ (i.e., VDL with highest bandwidth demand) and tries to place it (along with its VM and DB, and all NTPP VLs) in a PDL among the feasible PDLs so that the total network cost incurred due to the placement is minimum [lines 2--29] (Figure \ref{chap4-fig-vdl-vcl-placement-1}(a)). As explained in Section \ref{chap4-sec-vl-place-feas}, there can be three cases for this placement depending on current placement status of the VDL's VM and DB. 

\begin{algorithm}[!t]
\scriptsize
\textbf{Input:} $DC$ and $AE$. \\
\textbf{Output:} Total network cost of $AE$ placement.
\begin{algorithmic}[1]
\STATE  $totCost \leftarrow 0;$ Sort $vdlList$ and $vclList$ in decreasing order of VL's bandwidth demands;

\WHILE{$vdlList \neq \emptyset$} \COMMENT{NDAP tries to place all VDLs in vdlList}
	\STATE $VDL \leftarrow vdlList[0]; minCost \leftarrow \infty; VM \leftarrow VDL.VM; DB \leftarrow VDL.DB; selCN \leftarrow null; selSN \leftarrow null;$

	\IF{$DN(VM) = null \wedge DN(DB) = null$} \COMMENT{Case 1.1: Both VM and DB are not placed}
		\FOR{each $CN \in cnList \wedge VMPeerFeas(VM, CN) = 1$}
			\FOR{each $SN \in snList \wedge DBPeerFeas(DB, SN) = 1$}
				\IF{$BW(VM, DB) \leq BA(CN,SN)$}
					\STATE $cost \leftarrow BW(VM,DB) \times DS(CN,SN) + VMPeerCost(VM,CN)+ DBPeerCost(DB,SN);$
					\STATE \textbf{if} $cost < minCost$ \textbf{then} $minCost \leftarrow cost; selCN \leftarrow CN; selSN \leftarrow SN;$ \textbf{endif}
				\ENDIF
			\ENDFOR
		\ENDFOR
		\STATE \textbf{if} $minCost \neq \infty$ \textbf{then} $BA(selCN, selSN) \leftarrow BA(selCN, selSN) - BW(VM, DB);$ \textbf{endif}

	\ELSIF{$DN(VM) = null \wedge DN(DB) \neq null$} \COMMENT{Case 1.2: VM is not placed and DB is already placed}
		\FOR{each $CN \in cnList \wedge VMPeerFeas(VM, CN) = 1$}
			\STATE $cost \leftarrow VMPeerCost(VM,CN);$
			\STATE \textbf{if} $cost < minCost$ \textbf{then} $minCost \leftarrow cost; selCN \leftarrow CN;$ \textbf{endif}
		\ENDFOR

	\ELSIF{$DN(VM) \neq null \wedge DN(DB) = null$} \COMMENT{Case 1.3: VM is already placed and DB is not placed}
		\FOR{each $SN \in snList \wedge DBPeerFeas(DB, SN) = 1$}
			\STATE $cost \leftarrow DBPeerCost(DB,SN);$
			\STATE \textbf{if} $cost < minCost$ \textbf{then} $minCost \leftarrow cost; selSN \leftarrow SN;$ \textbf{endif}
		\ENDFOR	
	\ENDIF
	
	\STATE \textbf{if} $minCost = \infty$ \textbf{then return} $-1;$ \textbf{endif} \COMMENT{Feasible placement not found}
	\STATE \textbf{if} $selCN \neq null$ \textbf{then} $PlaceVMandPeerVLs(VM, selCN);$ \textbf{endif} \COMMENT{For Case 1.1 and Case 1.2}
	\STATE \textbf{if} $selSN \neq null$ \textbf{then} $PlaceDBandPeerVLs(DB, selSN);$ \textbf{endif} \COMMENT{For Case 1.1 and Case 1.3}
	\STATE $totCost \leftarrow totCost + minCost; vdlList.remove(0);$
\ENDWHILE

\WHILE{$vclList \neq \emptyset$} \COMMENT{NDAP tries to place remaining VCLs in vclList}
	\STATE $VCL \leftarrow vclList[0]; minCost \leftarrow \infty; VM_{1} \leftarrow VCL.VM_{1}; VM_{2} \leftarrow  VCL.VM_{2}; selCN_{1} \leftarrow null; selCN_{2} \leftarrow null;$

	\IF{$DN(VM_{1}) = null \wedge DN(VM_{2}) = null$} \COMMENT{Case 2.1: Both VMs are not placed}
		\FOR{each $CN_{1} \in cnList \wedge VMPeerFeas(VM_{1}, CN_{1}) = 1$}
			\FOR{each $CN_{2} \in cnList \wedge VMPeerFeas(VM_{2}, CN_{2}) = 1$}
				\STATE \textbf{if} $CN_{1} = CN_{2} \wedge VMPeerFeas(VM_{1},VM_{2},CN) = 0$ \textbf{then continue}; \textbf{endif}
				\IF{$BW(VM_{1}, VM_{2}) \leq BA(CN_{1},CN_{2})$}
					\STATE $cost \leftarrow BW(VM_{1},VM_{2}) \times DS(CN_{1},CN_{2});$
					\STATE $cost \leftarrow cost + VMPeerCost(VM_{1},CN_{1})+ VMPeerCost(VM_{2}, CN_{2});$
					\STATE \textbf{if} $cost < minCost$ \textbf{then} $minCost \leftarrow cost; selCN_{1} \leftarrow CN_{1}; selCN_{2} \leftarrow CN_{2};$ \textbf{endif}
				\ENDIF
			\ENDFOR
		\ENDFOR
		\STATE \textbf{if} $minCost \neq \infty$ \textbf{then} $BA(selCN_{1}, selCN_{2}) \leftarrow BA(selCN_{1}, selCN_{2}) - BW(VM_{1}, VM_{2});$ \textbf{endif}

	\ELSIF{$DN(VM_{1}) \neq null \vee DN(VM_{2}) \neq null$} \COMMENT{Case 2.2: One of the VMs is not placed}
		\STATE \textbf{if} $DN(VM_{1}) \neq null$ \textbf{then} $\text{swap values of }VM_{1} \text{ and }VM_{2};$ \textbf{endif} \COMMENT{Now $VM_{1}$ denotes the not-yet-placed VM}
		\FOR{each $CN_{1} \in cnList \wedge VMPeerFeas(VM_{1}, CN_{1}) = 1$}
			\STATE $cost \leftarrow VMPeerCost(VM_{1},CN_{1});$
			\STATE \textbf{if} $cost < minCost$ \textbf{then} $minCost \leftarrow cost; selCN_{1} \leftarrow CN_{1};$ \textbf{endif}
		\ENDFOR
	\ENDIF

	\STATE \textbf{if} $minCost = \infty$ \textbf{then return} $-1;$ \textbf{endif} \COMMENT{Feasible placement not found}

	\STATE	$PlaceVMandPeerVLs(VM_{1}, selCN_{1});$ \COMMENT{For Case 2.1 and Case 2.2}
	\STATE \textbf{if} $selCN_{2} \neq null$ \textbf{then} $PlaceVMandPeerVLs(VM_{2}, selCN_{2});$ \textbf{endif} \COMMENT{For Case 2.1}
	
	\STATE $totCost \leftarrow totCost + minCost; vclList.remove(0);$
	
\ENDWHILE
\STATE \textbf{return} $totCost;$

\end{algorithmic}
\caption{NDAP Algorithm}
\label{chap4-alg-ndap-1}
\end{algorithm}

When the VDL matches Case 1.1 (both VM and DB are not placed), then for each feasible CN and SN in DC (Eq. \ref{chap4-eq-vm-place-feas-1} and \ref{chap4-eq-db-place-feas-1}), it is checked if the bandwidth demand of the VDL can be satisfied by the available bandwidth of the corresponding PDL connecting the CN and SN. If it can be satisfied, then the total cost of placing the VDL and its associated NTPP VLs is measured (Eq. \ref{chap4-eq-vm-place-cost-1} and \ref{chap4-eq-db-place-cost-1}). The $\langle CN,SN \rangle$ pair that offers the minimum cost is selected for placing the $\langle VM,DB \rangle$ pair and the available bandwidth capacity of the PDL that connects the selected $\langle CN,SN \rangle$ pair is updated to reflect the VDL placement [lines 4--13]. When the VDL matches Case 1.2 (VM is not placed, but DB is placed), the feasible CN that offers minimum cost placement is selected for the VM and the total cost is measured [lines 14--18]. In a similar way, Case 1.3 (VM is placed, but DB is not placed) is handled in lines 19--24 and the best SN is selected for the DB placement. 

If NDAP fails to find a feasible CN or SN, it returns $-1$ to indicate failure in finding a feasible placement for the AE [line 25]. Otherwise, it activates the placements of the VM and DB along with their NTPP VLs by using subroutines $PlaceVMandPeerVLs$ (Algorithm \ref{chap4-alg-placeVMandpeerVLs-1}) and $PlaceDBandPeerVLs$ (Algorithm \ref{chap4-alg-placeDBandpeerVLs-1}), accumulates the measured cost in variable $totCost$, and removes the VDL from $vdlList$ [lines 26--28]. In this way, by picking the VDLs from a list that is already sorted based on bandwidth demand and trying to place each VDL, along with its NTPP VLs, in such a way that the incurred network cost is minimum in the current context of the DC resource state, NDAP strives to minimize the total network cost of placing the AE as formulated by the objective function $f$ (Eq. \ref{chap4-eq-ndap-obj-func-1}) of the proposed optimization. In particular, in each iteration of the first while loop (lines 2--29), NDAP pick the next highest bandwidth demanding VDL from the $vdlList$ and finds the best placement (i.e., minimum cost) for it along with its NTPP VLs. Moreover, the placement of the VDLs are performed before the placement of the VCLs since the average VDL bandwidth demand is expected to be higher than the average VCL bandwidth demand considering the fact that the average traffic volume for $\langle VM,DB \rangle$ pair is supposed to be higher than that for $\langle VM,VM \rangle$ pair.

After NDAP has successfully placed all the VDLs, then it starts placing the remaining VCLs in the $vclList$ (i.e., VCLs that were not NTPP VLs during the VDLs placement). For this part of the placement, NDAP applies a similar approach by repeatedly taking the first VCL from $vclList$ and trying to place it on a feasible PCL so that the incurred network cost is minimum [lines 30--55] (Figure \ref{chap4-fig-vdl-vcl-placement-1}(b)). This time, there can be two cases depending on the placement status of the two VMs of the VCL (Section \ref{chap4-sec-vl-place-feas}). 

When the VCL matches Case 2.1 (both VMs are not placed), then for each feasible CN in DC (Eq. \ref{chap4-eq-vm-place-feas-1}), it is first checked if both the VMs ($VM_{1}$ and $VM_{2}$) are being tried for placement in the same CN. In such cases, if the combined placement of both the VMs along with their NTPP VLs are not feasible (Eq. \ref{chap4-eq-vm-place-feas-2}), then NDAP continues checking feasibility for different CNs [line 35]. When both VMs placement feasibility passes and the bandwidth demand of the VCL can be satisfied by the available bandwidth of the corresponding PCL connecting the CNs, then the total cost of placing the VCL and its associated NTPP VLs is measured (Eq. \ref{chap4-eq-vm-place-cost-1} and \ref{chap4-eq-db-place-cost-1}) [lines 36--40]. When both the VMs are being tried for the same CN, then they can communicate with each other using memory copy rather going through physical network link and the available bandwidth check in line 36 works correctly since the intra-CN available bandwidth is considered to be unlimited. The $\langle CN_{1}, CN_{2} \rangle$ pair that offers the minimum cost is selected for placing the $\langle VM_{1},VM_{2} \rangle$ pair and the available bandwidth capacity of the PCL connecting the selected $\langle CN_{1}, CN_{2} \rangle$ pair is updated to reflect the VCL placement [lines 39--43]. When the VCL matches Case 2.2 (one of the VMs is not placed), the feasible CN that offers minimum cost placement is selected for the not-yet-placed VM ($VM_{1}$) and the total cost is measured [lines 44--50].

Similar to VDL placement, if NDAP fails to find feasible CNs for any VCL placement, it returns $-1$ to indicate failure [line 51]. Otherwise, it activates the placements of the VMs along with their NTPP VLs by using subroutine $PlaceVMandPeerVLs$ (Algorithm \ref{chap4-alg-placeVMandpeerVLs-1}), accumulates the measured cost in $totCost$, and removes the VCL from $vclList$ [lines 52--55]. For the same reason as for VDL placement, the VCL placement part of the NDAP algorithm fosters the reduction of the objective function $f$ value (Eq. \ref{chap4-eq-ndap-obj-func-1}).

Finally, NDAP returns the total cost of the $AE$ placement, which also indicates a successful placement [line 56].

\section{Performance Evaluation}
\label{chap4-sec-perf-eval}
This section describes the performance of the proposed NDAP algorithm compared to other algorithms through a set of simulation based experiments. Section \ref{chap4-sec-alg-impl} gives a brief description of the evaluated algorithms, Section \ref{chap4-sec-sim-setup} describes the various aspects of the simulation environment, and finally, the results are discussed in the subsequent sections.

\subsection{Algorithms Compared}
\label{chap4-sec-alg-impl}
The following algorithms are evaluated and compared in this work:

\textit{Network-aware VM Allocation (NVA):} This is an extended version of the network-aware VM placement approach proposed by \citet{Piao2010} where the authors have considered already-placed data blocks. In this version, each $DB \in DBS$ is placed randomly in a $SN \in SNS$. Afterwards, each $VM$ that has one or more VDL is placed according to the VM allocation algorithm presented by the authors, provided that all of its NTPP VLs are placed on feasible PLs. For any remaining $VM \in VMS$, it is placed randomly. All the above placements are subject to the constraints presented in Eq. \ref{chap4-eq-cn-cpu-constr-1}, \ref{chap4-eq-cn-mem-constr-1}, \ref{chap4-eq-sn-str-constr-1}, \ref{chap4-eq-cn-cn-ba-constr-1}, and \ref{chap4-eq-cn-sn-ba-constr-1}. In order to increase the probability of feasible placements, DB and VM placements are tried multiple times and the maximum number of tries ($N_{mt}$) is parameterized by a constant which is set to 100 in the simulation. For the above mentioned implementation, the worst-case time complexity of NVA algorithm is given by:
\begin{equation}
\label{chap4-eq-nva-time-complx-1}
\begin{aligned}
T_{NVA} = \mathcal{O}(N_{d}N_{mt}) + \mathcal{O}(N_{v}N_{c}N_{vn}) + \mathcal{O}(N_{v}N_{mt}).
\end{aligned}
\end{equation}
Given the fact that $N_{mt}$ is a constant and the maximum number of VMs ($N_{v}$) and DBs ($N_{d}$) in an AE is generally much less than the number of computing nodes ($N_{c}$) in DC, the above time complexity reduces to:
\begin{equation}
\label{chap4-eq-nva-time-complx-2}
\begin{aligned}
T_{NVA} = \mathcal{O}(N_{v}N_{c}N_{vn}).
\end{aligned}
\end{equation}
Given that NVA starts with already-placed DBs, and VM placements are done in-place using no auxiliary data structure, NVA algorithm itself does not have any memory overhead.

\textit{First Fit Decreasing (FFD):} This algorithm begins by sorting the CNs in $cnList$ and SNs in $snList$ in decreasing order based on their remaining resource capacities. Since, CNs have two different types of resource capacities (CPU and memory), L1-norm mean estimator is used to convert the vector representation of multi-dimensional resource into scalar form. Similarly, all the VMs in $vmList$ and DBs in $dbList$ are sorted in decreasing order of their resource demands, respectively. Then, FFD places each DB from $dbList$ in the first feasible SN of $snList$ according to the \textit{First First} (FF) algorithm. Afterwards, it places each VM from $vmList$ in the first feasible CN of $cnList$ along with any associated NTPP VLs. All the above placements are subject to the constraints presented in Eq. \ref{chap4-eq-cn-cpu-constr-1}, \ref{chap4-eq-cn-mem-constr-1}, \ref{chap4-eq-sn-str-constr-1}, \ref{chap4-eq-cn-cn-ba-constr-1}, and \ref{chap4-eq-cn-sn-ba-constr-1}. For this implementation of FFD, the worst-case time complexity of FFD algorithm is given by:
\begin{equation}
\label{chap4-eq-ffd-time-complx-1}
\begin{aligned}
T_{FFD} = \mathcal{O}(N_{c}\text{lg}N_{c}) + \mathcal{O}(N_{s}\text{lg}N_{s}) + \mathcal{O}(N_{v}\text{lg}N_{v}) + \mathcal{O}(N_{d}\text{lg}N_{d}) + \mathcal{O}(N_{d}N_{s}) + \mathcal{O}(N_{v}N_{c}).
\end{aligned}
\end{equation}
Given the fact that, in a typical setting the number of VMs ($N_{v}$) and DBs ($N_{d}$) in an AE is much less than the number of CNs ($N_{c}$) and SNs ($N_{s}$) in DC, respectively, the above term reduces to:
\begin{equation}
\label{chap4-eq-ffd-time-complx-2}
\begin{aligned}
T_{FFD} = \mathcal{O}(N_{c}\text{lg}N_{c}) + \mathcal{O}(N_{s}\text{lg}N_{s}) + \mathcal{O}(N_{d}N_{s}) + \mathcal{O}(N_{v}N_{c}).
\end{aligned}
\end{equation}
Given that merge sort \citep{Cormen2001} is used in FFD to sort $cnList$, $snList$, $vmList$, and $dbList$, and $N_{c}$ is usually greater than each of $N_{s}$, $N_{v}$, and $N_{d}$ in a typical setting, it can be concluded that the memory overhead for the sorting operation is $\mathcal{O}(N_{c})$. Apart from sorting, the placement decision part of FFD works in-place without using any additional data structure. Therefore, the memory overhead of FFD algorithm is $\mathcal{O}(N_{c})$.

\textit{Network- and Data-aware Application Placement (NDAP):} The NDAP algorithm is implemented primarily based on the the description presented in Section \ref{chap4-sec-proposed-solution} and follows the execution flow presented in Algorithm \ref{chap4-alg-ndap-1}. The final NDAP algorithm utilizes the feasibility check (Eq. \ref{chap4-eq-vm-place-feas-1}, \ref{chap4-eq-vm-place-feas-2}, and \ref{chap4-eq-db-place-feas-1}), network cost computation (Eq. \ref{chap4-eq-vm-place-cost-1} and \ref{chap4-eq-db-place-cost-1}), and the placement subroutines (Algorithm \ref{chap4-alg-placeVMandpeerVLs-1} and \ref{chap4-alg-placeDBandpeerVLs-1}). All of these NDAP components need to go through a list of NTPP VLs for the corresponding VM or DB, and in the implementation, this list is stored in an array. Thus, the time complexity for each of these NDAP components is $\mathcal{O}(N_{vn})$. For the above mentioned implementation, the running time of NDAP algorithm (refering to peudocode in Algorithm \ref{chap4-alg-ndap-1}) is the sum of the time needed for sorting $vdlList$ and $vdlList$ ($T_{2}$), the time needed for placing all the VDLs in $vdlList$ ($T_{3\text{-}34}$), and the time needed for placing all the remaining VCLs in $vclList$ ($T_{36\text{-}65}$). 
The time complexity for placing a single VDL (considering three cases) is given by: 
\begin{equation}
\label{chap4-eq-ndap-time-complx-1}
\begin{aligned}
T_{6\text{-}33} &= \mathcal{O}(N_{c}N_{s}N_{vn}) + \mathcal{O}(N_{c}N_{vn}) + \mathcal{O}(N_{s}N_{vn}) + \mathcal{O}(N_{vn}) \\
				&= \mathcal{O}(N_{c}N_{s}N_{vn}).
\end{aligned}
\end{equation}
Therefore, the time complexity for placing all the VDLs is:
\begin{equation}
\label{chap4-eq-ndap-time-complx-2}
\begin{aligned}
T_{3\text{-}34} &= \mathcal{O}(N_{vd}N_{c}N_{s}N_{vn}).
\end{aligned}
\end{equation}
Similarly, the time complexity for placing all the remaining VCLs is:
\begin{equation}
\label{chap4-eq-ndap-time-complx-3}
\begin{aligned}
T_{36\text{-}65} &= \mathcal{O}(N_{vc}N_{c}^{2}N_{vn}).
\end{aligned}
\end{equation}
Thus, the worst-case time complexity of NDAP algorithm is given by:
\begin{equation}
\label{chap4-eq-ndap-time-complx-4}
\begin{aligned}
T_{NDAP} &= T_{2} + T_{3\text{-}34} + T_{36\text{-}65} \\
		 &= \mathcal{O}(N_{vd}\text{lg}N_{vd}) + \mathcal{O}(N_{vc}\text{lg}N_{vc}) + \mathcal{O}(N_{vd}N_{c}N_{s}N_{vn}) + \mathcal{O}(N_{vc}N_{c}^{2}N_{vn}).
\end{aligned}
\end{equation}
For this implementation of NDAP algorithm, merge sort is used in order to sort $vdlList$ and $vclList$ [line 2, Algorithms \ref{chap4-alg-ndap-1}]. Given that AEs are typically constituted of a number of VMs and DBs with sparse communication links between them, it is assumed that $N_{vd} = N_{vc} = \mathcal{O}(N_{v})$ since $N_{vd}$ and $N_{vc}$ are of the same order. Thus, the memory overhead for this sorting operation is $\mathcal{O}(N_{v})$. Apart from sorting, the placement decision part of NDAP [lines 3--67] works in-place and no additional data structure is needed. Therefore, the memory overhead of NDAP algorithm is $\mathcal{O}(N_{v})$.

The detailed computational time complexity analyses presented above may be further simplified as follows. While the number of computing node outweighs the number of storage node in a typical DC, these may be assumed of the same order, i.e., $N_{s} = \mathcal{O}(N_{c})$. Moreover, the size of a typical DC is at least multiple order higher than that of an AE. Hence, it can also be assumed that $N_{v}, N_{d}, N_{vc}, N_{vd}, N_{vn} = o(N_{c})$. From Eq. \ref{chap4-eq-nva-time-complx-2}, \ref{chap4-eq-ffd-time-complx-2}, \& \ref{chap4-eq-ndap-time-complx-4}, it can be concluded that the running time of NVA, FFD, and NDAP algorithms are $\mathcal{O}(N_{c})$, $\mathcal{O}(N_{c}\text{lg}N_{c})$, and $\mathcal{O}(N_{c}^{2})$, respectively, i.e., these are linear, linearithmic, and quadratic time algorithms, respectively. Regarding the overhead of the above mentioned algorithms, although there are variations in the run-time memory overhead, considering that the input optimization problem (i.e., AE placement in DC) itself has $\mathcal{O}(N_{c})$ memory overhead, it can be concluded that, overall, all the compared algorithms have equal memory overhead of $\mathcal{O}(N_{c})$.

For all the above algorithms, if any feasible placement is not found for a VM or DB, the corresponding algorithm terminates with failure status. The algorithms are implemented in Java (JDK and JRE version 1.7.0) and the simulation is conducted on a Dell Workstation (Intel Core i5-2400 3.10 GHz CPU (4 cores), 4 GB of RAM, and 240 GB storage) hosting Windows 7 Professional Edition.

\subsection{Simulation Setup}
\label{chap4-sec-sim-setup}

\subsubsection{Data Center Setup}
\label{chap4-sec-data-center-setup}
\begin{figure}[t]
\centering
\includegraphics[scale=0.4, trim=1cm 0cm 0cm 0cm]{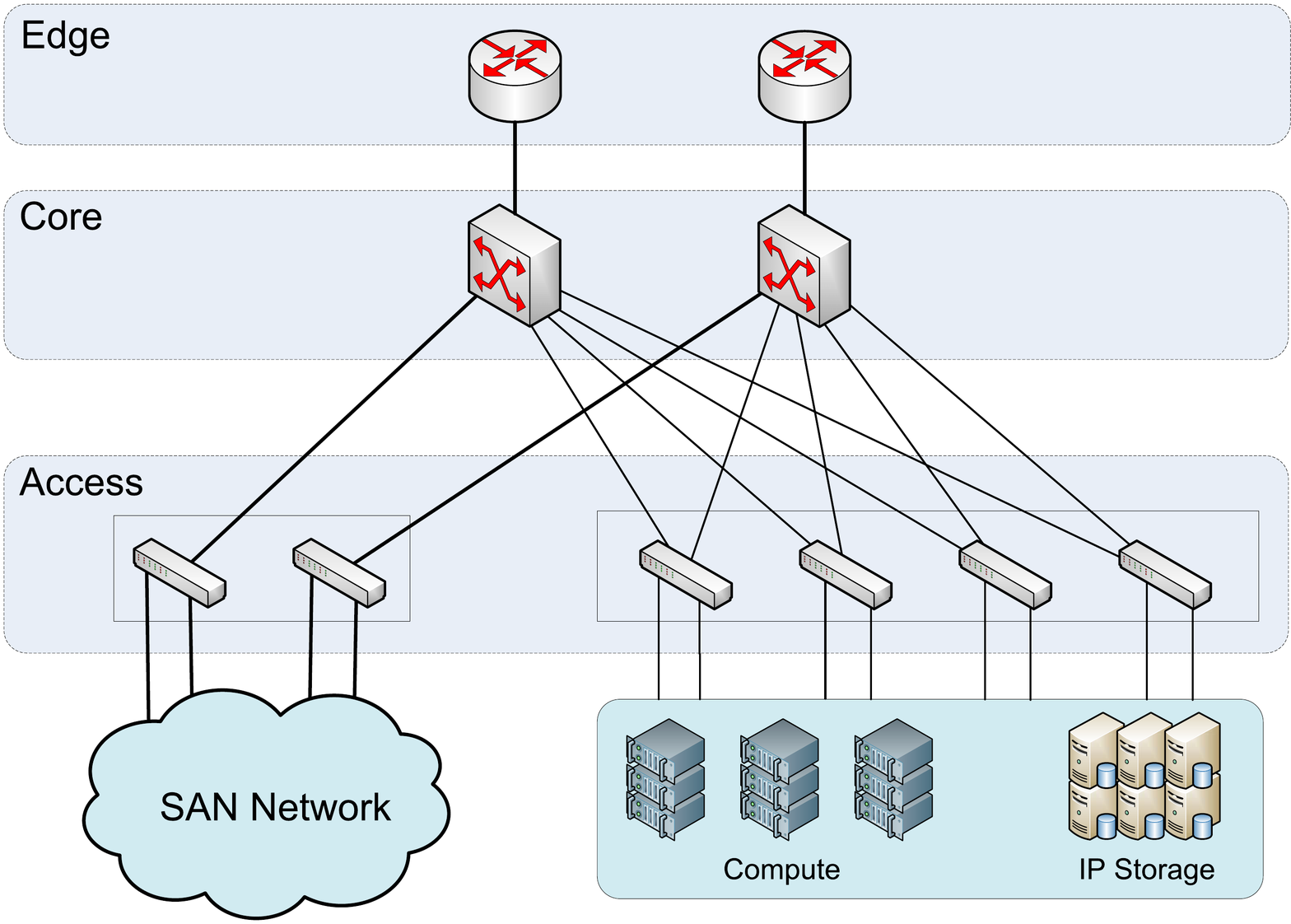}
\caption{Cloud-ready data center network architecture.}
\label{chap4-fig-cloud-ready-data-center-network-1}
\end{figure}

In order to address the increasing complexity of large-scale Cloud data centers, network vendors are coming up with network architecture models focusing on the resource usage patterns of Cloud applications. For example, Juniper Networks Inc. in their "Cloud-ready data center reference architecture" suggests the use of Storage Area Networks (SAN) interconnected to the computing network with converged access switches \citep{Juniper2012}, similar to the one shown in Figure \ref{chap4-fig-cloud-ready-data-center-network-1}. The simulated data center is generated following this reference architecture with three-tier computing network topology (core-aggregation-access) \citep{Kliazovich2013} and SAN-based storage network. Following the approach presented in \citep{Korupolu2009}, the number of parameters is limited in simulating the data center by using the number of physical computing servers as the only parameter denoted by $N$. The number of other data center nodes are derived from $N$ as follows: $5N/36$ high-end storage devices with built-in spare computing resources that work as multi-function devices for storage and computing, $4N/36 (= N/9)$ regular storage devices without additional computing resources, $N/36$ high-end core switches with built-in spare computing resources that work as multi-function devices for switching and computing, $N/18$ mid-level aggregation switches, and $5N/12\text{ }(= N/3 + N/12)$ access switches. Following the three-tier network topology \citep{Kliazovich2013}, $N/3$ access switches provide connectivity between $N$ computing servers and $N/18$ aggregation switches, whereas the $N/18$ aggregation switches connects $N/3$ access switches and $N/36$ core switches in the computing network. The remaining $N/12$ access switches provide connectivity between $N/4$ storage devices and $N/36$ core switches in the storage network. In such a data center setup, the total number of computing nodes (CNs) $N_{c} = N + 5N/36 + N/36 = 7N/6$ and the total number of storage nodes (SNs) $N_{s} = 5N/36 + 4N/36 = N/4$.

Network distance between $\langle CN,CN \rangle$ pairs and between $\langle CN,SN \rangle$ pairs are measured as $DS = h \times DF$, where $h$ is the number of physical hops between two DC nodes (CN or SN) in the simulated data center architecture as defined above, and $DF$ is the \textit{Distance Factor} that implies the physical inter-hop distance. The value of $h$ is computed using the analytical expression for tree topology as presented in \citep{Meng2010} and $DF$ is fed as a parameter to the simulation. Network distance of a node with itself is $0$ which implies that data communication is done using memory copy without going through the network. A higher value of $DF$ indicates greater relative communication distance between any two data center nodes.

\subsubsection{Application Environment Setup}
\label{chap4-sec-app-env-setup}
\begin{figure}[t]
\centering
\includegraphics[scale=0.6, trim=1.5cm 4cm 0cm 2cm]{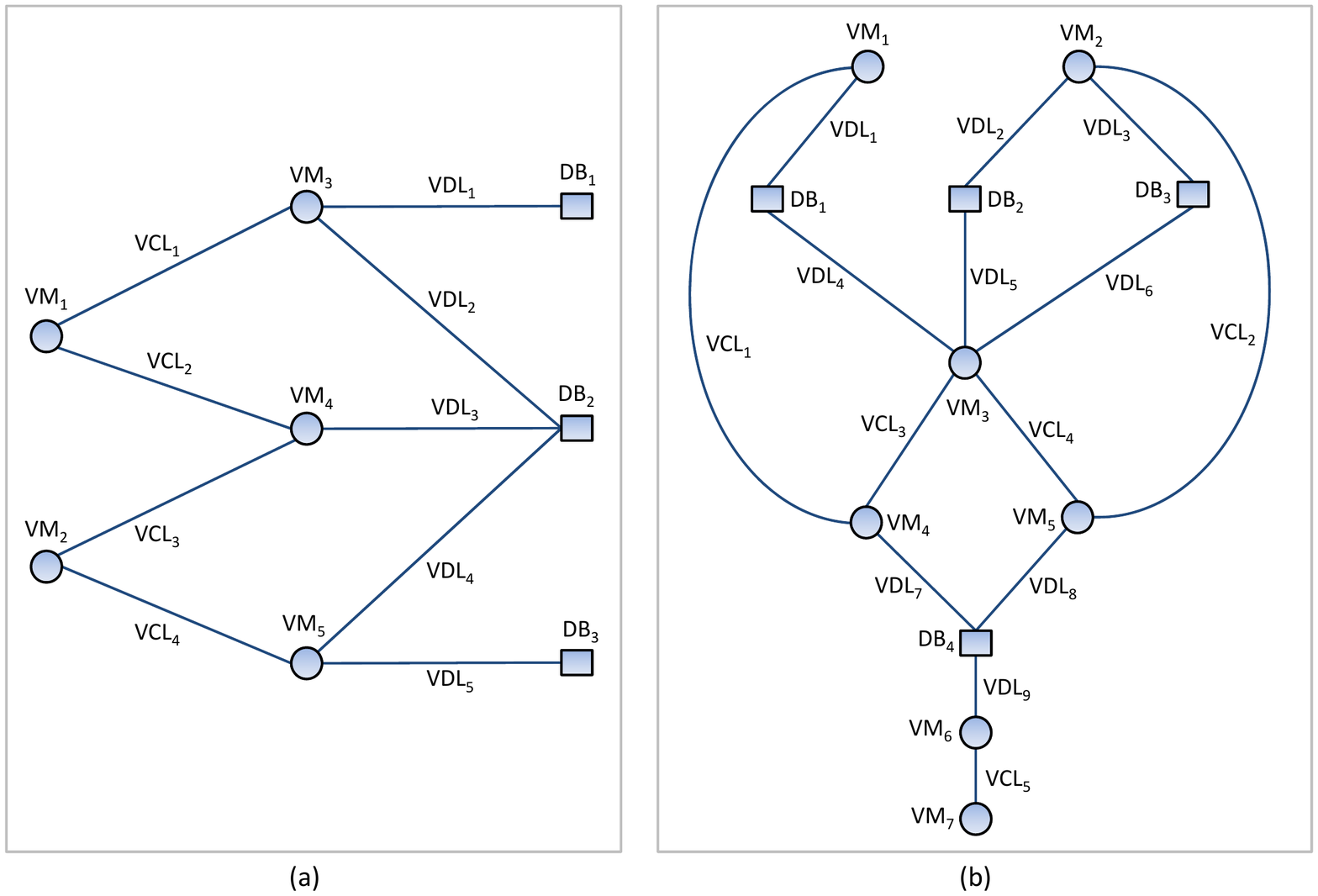}
\caption{Application environment models for (a) Multi-tier application and (b) Scientific (Montage) workflow.}
\label{chap4-fig-ae-models-1}
\end{figure}

In order to model composite application environments for the simulation, multi-tier enterprise applications and scientific workflows are considered as representatives of the dominant Cloud applications. According to the analytical model for multi-tier Internet applications presented in \citep{Urgaonkar2005}, three-tier applications are modeled as comprised of 5 VMs ($N_{v}=5$) and 3 DBs ($N_{d}=3$) interconnected through 4 VCLs ($N_{vc}=4$) and 5 VDLs ($N_{vd}=5$) as shown in Figure \ref{chap4-fig-ae-models-1}(a). In order to model scientific applications, Montage workflow is simulated as composed of 7 VMs ($N_{v}=7$) and 4 DBs ($N_{d}=4$) interconnected through 5 VCLs ($N_{vc}=5$) and 9 VDLs ($N_{vd}=9$) following the structure presented in \citep{Juve2013} (Figure \ref{chap4-fig-ae-models-1}(b)). While deploying an application in data center, user provided hints on estimated resource demands are parameterized during the course of the experimentation. Extending the approaches presented in \citet{Meng2010} and in \citet{Shrivastava2011}, computing resource demands (CPU and memory) for VMs, storage resource demands for DBs, and bandwidth demands for VLs are stochastically generated based on normal distribution with parameter means ($meanCom$, $meanStr$, and $meanVLBW$, respectively) and standard deviation ($sd$) against normalized total resource capacities of CNs and SNs, and bandwidth capacities of PLs, respectively.

\color{black}
\subsubsection{Simulated Scenarios}
\label{chap4-sec-sim-scen}
For each of the experiments, all the algorithms start with their own empty data centers. In order to represent the dynamics of the real Cloud data centers, two types of events are simulated: (1) AE deployment and (2) AE termination. With the purpose of assessing the relative performance of the various placement algorithms in states of both higher and lower resource availability of data center nodes (CNs and SNs) and physical links (PCLs and PDLs), the experiments simulated scenarios where the average number of AE deployments doubles the average number of AE terminations. Since during the initial phase of the experiments the data centers are empty, algorithms enjoy more freedom for the placement of AE components. Gradually, the data centers get loaded due to higher number of AE deployments compared to the number of AE terminations. In order to reflect upon the reality of application deployment dynamics in real Clouds where the majority of the Cloud application spectrum is composed of multi-tier enterprise applications, in the simulated scenarios, 80\% of the AE deployments are considered to be enterprise applications (three-tier application models) and 20\% are considered as scientific applications (Montage workflow models). Overall, the following two scenarios are considered:

\textit{Group Scenario:} For all the placement algorithms, AE deployments and terminations are continued until any of them fails to place an AE due to the lack of feasible placement. For maintaining fairness among algorithms, the total number of AE deployments and terminations for each of the placement algorithms are equal and the same instances of AEs are deployed or terminated for each simulated event.

\textit{Individual Scenario:} For each of the algorithms, AE deployment and termination is continued separately until it fails to place an AE due to the lack of a finding feasible placement. Similar to the group scenario, all the algorithms draw AEs from same pools so that all the algorithms work with the same AE for each event.
\color{black}

All the experiments presented in this paper are repeated 1000 times and the average results are reported.

\subsubsection{Performance Evaluation Metrics}
\label{chap4-sec-perf-eval-metrics}
In order to assess the network load imposed due the placement decisions, the average network cost of AE deployment is computed (using objective function $f$ accordingly to Eq. \ref{chap4-eq-ndap-obj-func-1}) for each of the algorithms in the group scenario. Since the cost functions (Eq. \ref{chap4-eq-cost-1} and \ref{chap4-eq-cost-2}) are defined based on network distance between DC nodes and expected amount of traffic flow, it effectively provides measures of the network packet transfer delays, and imposed packet forwarding load and power consumption for the network devices (e.g., switches and routers) and communication links. With the aim of maintaining a fair comparison among the algorithms, the average cost metric is computed and compared in the group scenario where all the algorithms terminate when any of them fails to place an AE due to the feasible resource constraints (Eq. \ref{chap4-eq-cn-cpu-constr-1}, \ref{chap4-eq-cn-mem-constr-1}, \ref{chap4-eq-sn-str-constr-1}, \ref{chap4-eq-cn-cn-ba-constr-1}, and \ref{chap4-eq-cn-sn-ba-constr-1}) in DC and, as a consequence, each algorithm works with the same instances of AE at each deployment and termination event, and the average cost is computed over the same number of AEs.

In order to measure how effectively each of the algorithms utilizes the network bandwidth during AE placements, the total number of AE deployments in empty DC is measured until the data center saturates in the individual scenario. Through this performance metric, the effective capacity of the DC resources utilized by each of the placement algorithms is captured and compared. 

In order to assess how effectively the placement algorithms localize network traffic and, eventually, optimize network performance, the average network utilization of access, aggregation, and core switches are measured in the group scenario. In this part of the evaluation, the group scenario is chosen so that when any of the algorithms fail to place an AE, all the algorithms halt their placements with the purpose of keeping the total network loads imposed on the respective data centers for each of the algorithms remain same. This switch-level network usage assessment is performed through scaling the mean and standard deviation of the VLs' bandwidth demands.

Finally, the average placement decision computation time for AE deployment is measured for the individual scenario. Average placement decision time is an important performance metric to assess the efficacy of NDAP as an on-demand AE placement algorithm and its scalability across various factors. 

\color{black}

All the above performance metrics are measured against the following scaling factors: (1) DC size , (2) mean resource demands of VMs, DBs, and VLs, (3) diversification of workloads, and (4) network distance factor $DF$. The following subsections present the experimental results and analysis for each of the experiments conducted.

\subsection{Scaling Data Center Size}
\label{chap4-sec-scaling-dc-size}
In this part of the experiment, the placement quality of the algorithms with increasing size of the DC is evaluated and compared. As mentioned in Section \ref{chap4-sec-data-center-setup}, $N$ is used as the only parameter to denote DC size, and its minimum and maximum values are set to 72 and 4608, respectively, doubling at each subsequent simulation phase. Thus, in the largest DC there are a total of 5376 CNs and 1152 SNs. The other parameters $meanCom$, $meanStr$, $meanVLBW$, $sd$, and $DF$ are set to 0.3, 0.4, 0.35, 0.5, and 2, respectively.

\begin{figure}[t]
\centering
\includegraphics[scale=0.55, trim=0cm 32cm 1.2cm 1.5cm]{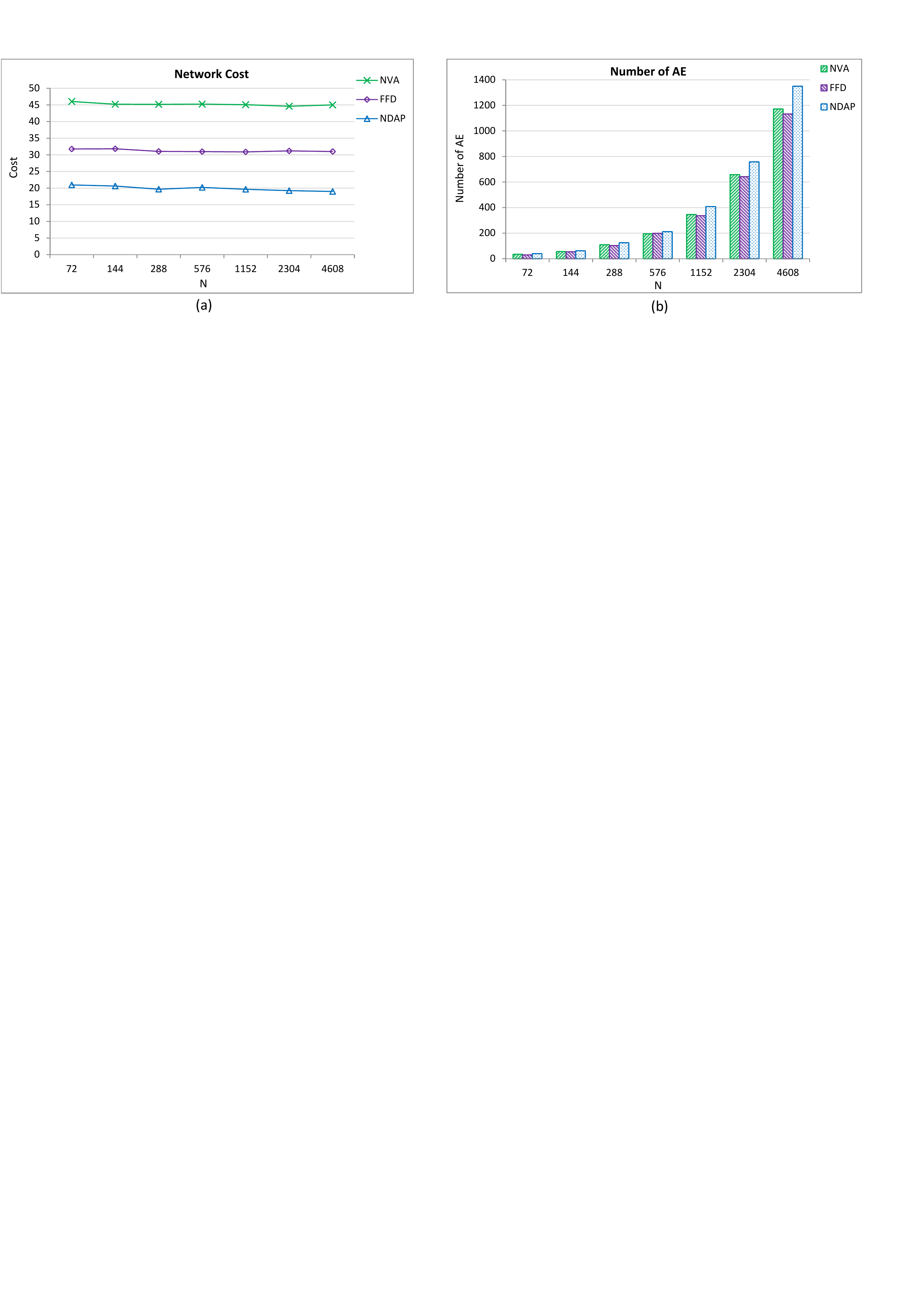}
\caption{Performance with increasing $N$: (a) Network cost and (b) Number of AE deployed in DC.}
\label{chap4-fig-perf-N}
\end{figure}

Figure \ref{chap4-fig-perf-N}(a) shows the average cost of AE placement incurred by each of the three algorithms in the group scenario for different values of $N$. From the chart, it is quite evident that NDAP consistently outperforms the other placement algorithms at a much higher level for the different DC sizes and its average AE placement cost is 56\% and 36\% less than NVA and FFD, respectively. Being network-aware, NDAP checks the feasible placements with the goal of minimizing the network cost. FFD, on the other hand, tries to place the ANs in DNs with maximum available resource capacities and, as a result, has possibility of placing VLs on shorter PLs. And, finally, NVA has random components in placement decisions and, thus, incurs higher average cost. 

From Figure \ref{chap4-fig-perf-N}(b), it can be observed that the average number of successful AE deployments in the individual scenario by the algorithms increases non-linearly with the DC size as more DNs and PLs (i.e., resources) are available for AE deployments. It is also evident that NDAP deploys larger number of AEs in data center compared to other algorithms until the data center is saturated with resource demands. The relative performance of NDAP remains almost steady across different data center sizes--- it deploys around 13-17\% and 18-21\% more AEs compared to NVA and FFD, respectively. This demonstrates the fact that NDAP's effectiveness in utilizing the data center resources is not affected by the scale of the data center.

\subsection{Variation of Mean Resource Demands}
\label{chap4-sec-var-mean-rsc-dem}
This experiment assesses the solution qualities of the placement algorithms when the mean resource demands of the AEs increase. Since the AE is composed of different components, the mean resource demands are varied in two different approaches presented in the rest of this subsection. As for the other parameters $N$, $sd$, and $DF$ are set to 1152, 0.4, and 2, respectively.

\subsubsection{Homogeneous Mean Resource Demands}
\label{chap4-sec-hom-mean-rsc-dem}
The same $mean$ (i.e., $meanCom=meanStr=meanVLBW=mean$) is used to generate the computing (CPU and memory) resource demands of VMs, storage resource demands of DBs, and bandwidth demands of VLs under normal distribution. The experiment starts with a small $mean$ of 0.1 and increases it upto 0.7, adding up with 0.1 at each subsequent phase.

\begin{figure}[t]
\centering
\includegraphics[scale=0.55, trim=0cm 32cm 2cm 1.5cm]{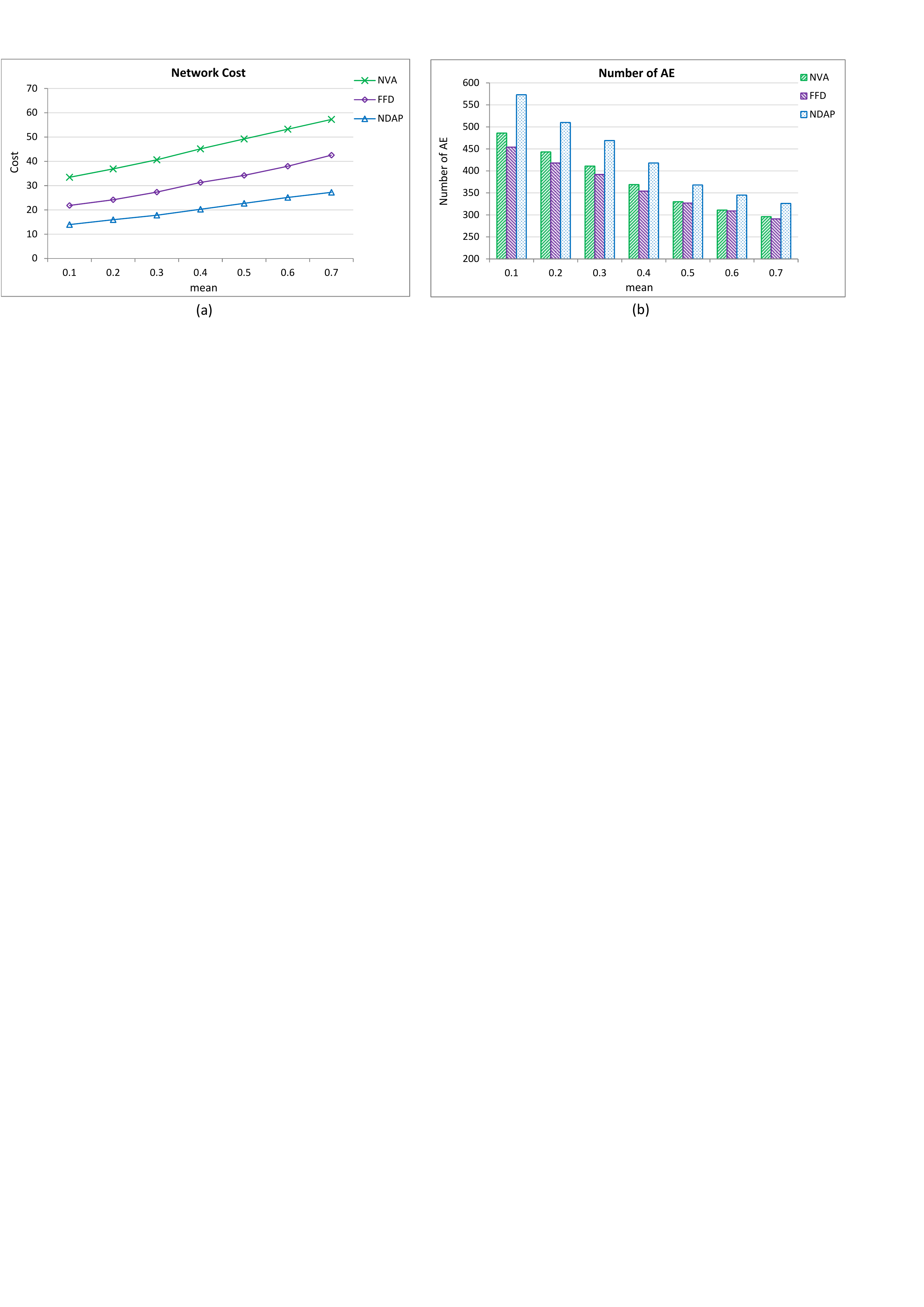}
\caption{Performance with increasing mean (homogeneous): (a) Network cost and (b) Number of AE deployed in DC.}
\label{chap4-fig-perf-mean-same}
\end{figure}

The average cost for AE placement is shown in Figure \ref{chap4-fig-perf-mean-same}(a) for the group scenario. It is obvious form the the chart that NDAP achieves much better performance compared to other placement algorithms--- on average it incurs 55\% and 35\% less cost compared to NVA and FFD, respectively. With the increase of mean resource demands, the incurred cost for each algorithm increases almost at a constant rate. The reason behind this performance pattern is that when the mean resource demands of the AE components (VMs, DBs, and VLs) increase with respect to the available resource capacities of the DC components (CNs, SNs, and PLs), the domain of feasible placements is reduced which causes the rise in the average network cost. 

Figure \ref{chap4-fig-perf-mean-same}(b) shows the average number of AEs deployed in empty DC with increasing $mean$ for the individual scenario. It can be seen from the chart that the number of AEs deployed by the algorithms constantly reduces as higher mean values are used to generate the resource demands. This is due to the fact that when resource demands are increased compared to the available resource capacities, the DC nodes and PLs can accommodate fewer number of AE nodes and VLs. One interesting observation from this figure is that FFD was able to deploy fewer number of AEs compared to NVA when the $mean$ was small. This can be attributed to the multiple random tries during ANs placement by NVA which helps it to find feasible placements, although with higher average cost. Overall, NDAP has been able to place larger number of AEs compared to other algorithms across all mean values: 10-18\% and 12-26\% more AEs than NVA and FFD, respectively.

\subsubsection{Heterogeneous Mean Resource Demands}
\label{chap4-sec-het-mean-rsc-dem}
In order to assess the performance variations across different mean levels of resource demands of AE components, two different mean levels $L$ (low) and $H$ (high) are set in this part of the experiment for mean VM computing resource demands ($meanCom$ for both CPU and memory), mean DB storage resource demands ($meanStr$), and mean VL bandwidth demands ($meanVLBW$). $L$ and $H$ levels are set to 0.2 and 0.7 for this simulation. Given the two levels for the three types of resource demands, there are eight possible combinations.

\begin{figure}[t]
\centering
\includegraphics[scale=0.59, trim=.8cm 32cm 0cm 1.5cm]{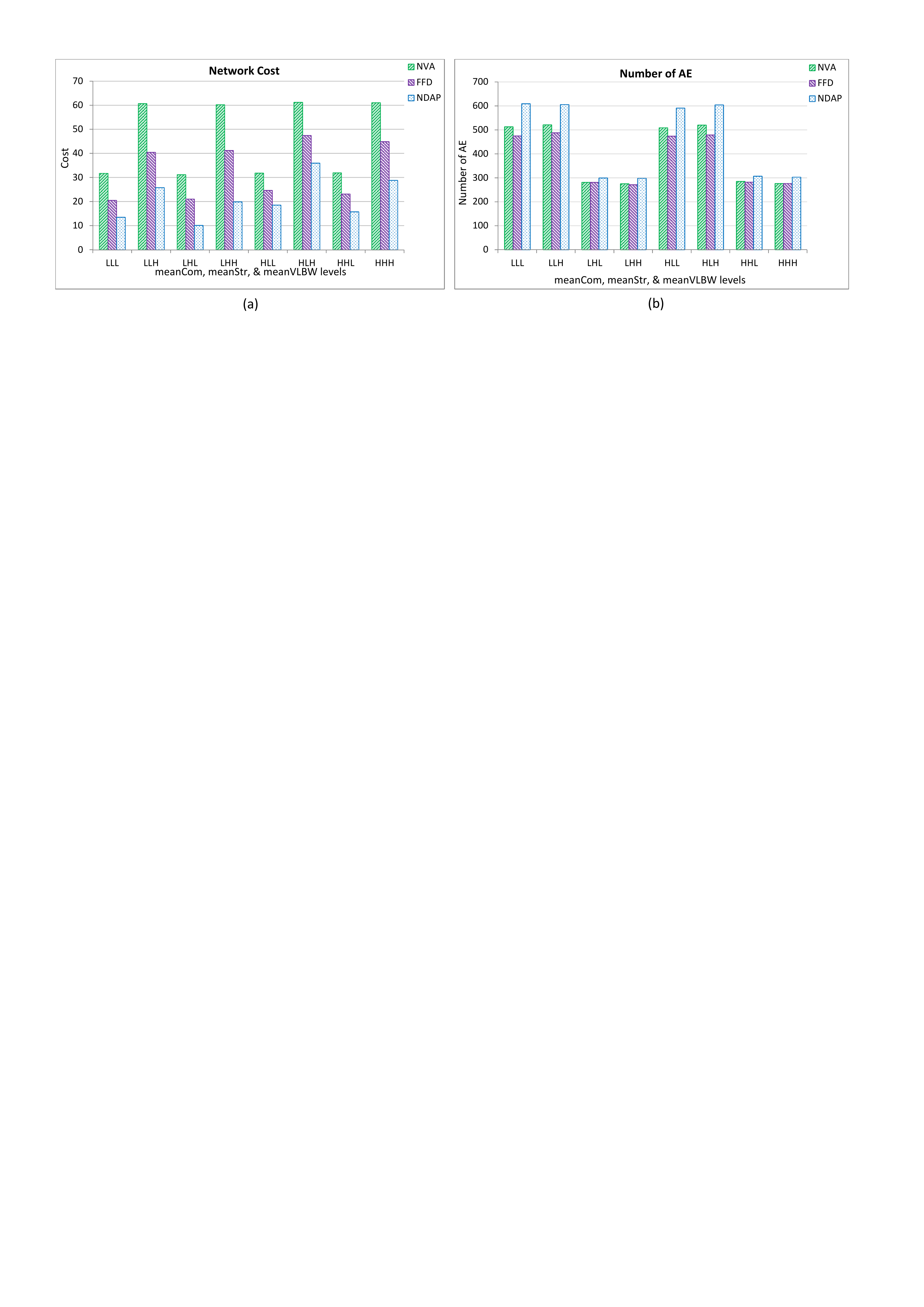}
\caption{Performance with mixed levels of means (heterogeneous): (a) Network cost and (b) Number of AE deployed in DC.}
\label{chap4-fig-perf-mean-mixed}
\end{figure}

Figure \ref{chap4-fig-perf-mean-mixed}(a) shows the average network costs of the three algorithms for the eight different mean levels ($x$ axis of the chart). The three different positions of the labels are set as follows: the left-most, the middle, and the right-most positions are for $meanCom$, $meanStr$, and $meanVLBW$, respectively. As the chart shows, NDAP performs much better in terms of incurred cost compared than the other algorithms for each of the mean combinations. Its relative performance is highest for combinations $LHL$ and $LHH$ incurring on average 67\% and 52\% less costs compared to NVA and FFD; whereas its performance is lowest for combinations $HLL$ and $HLH$ incurring on average 42\% and 25\% less costs compared to NVA and FFD, respectively. The reason behind this pattern is the algorithmic flow of NDAP as it starts with the VDLs placement and finishes with the remaining VCLs placement. As a consequence, for relatively higher mean of DB storage demands, NDAP relatively performs better.

A similar performance trait can be seem in Figure \ref{chap4-fig-perf-mean-mixed}(b) that shows that NDAP places more AEs in DC compared to other algorithms. An overall pattern demonstrated by the figure is that when the $meanStr$ is high ($H$), the number of AEs deployed is reduced for all algorithms compared to the cases when $meanStr$ is low ($L$). This is because the simulated storage resources are fewer compared to the computing and network resources of DC with respect to the storage, computing, and bandwidth demands of AEs, respectively. Since NDAP starts AE deployment with efficient placement of DBs and VDLs, on average it deploys 17\% and 26\% more AEs compared to NVA and FFD, respectively, when $meanStr=H$; whereas this improvement is 9\% for both NVA and FFD when $meanStr=L$.

\subsection{Diversification of Workloads}
\label{chap4-sec-div-workloads}
This part of the experiment simulates the degree of workload diversification of the deployed AEs through varying the standard deviation of the random (normal) number generator used to generate the resource demands of the components of AEs. For this purpose, the initial value for $sd$ parameter is set to 0.05 and increased gradually by adding 0.05 at each simulation phase until a maximum of 0.5 is reached. The other parameters $N$, $meanCom$, $meanStr$, $meanVLBW$, and $DF$ are set to 1152, 0.3, 0.4, 0.35, and 2, respectively.

\begin{figure}[t]
\centering
\includegraphics[scale=0.55, trim=0cm 32.5cm 0.4cm 2cm]{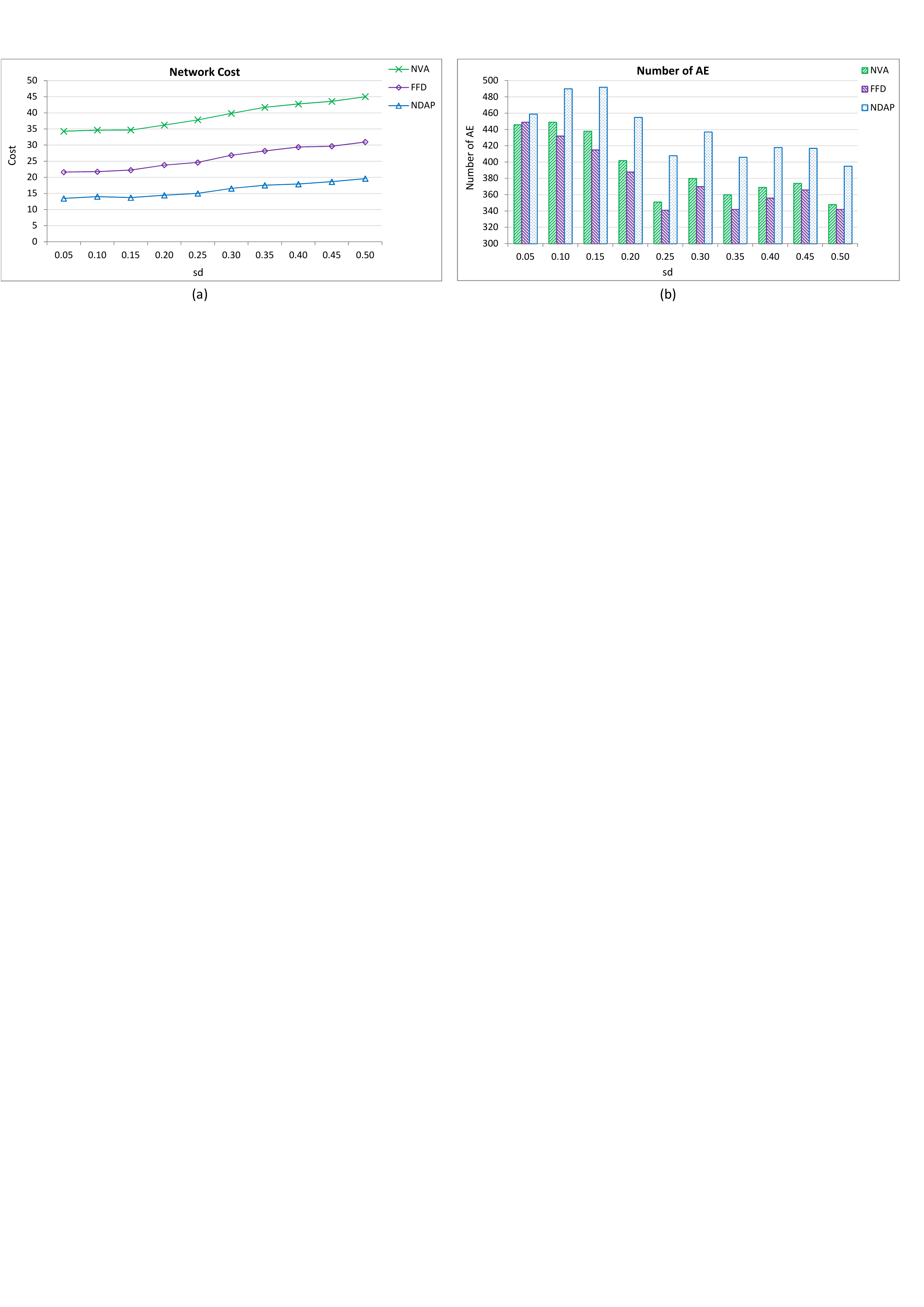}
\caption{Performance with increasing standard deviation of resource demands: (a) Network cost and (b) Number of AE deployed in DC.}
\label{chap4-fig-perf-sd}
\end{figure}

As shown in Figure \ref{chap4-fig-perf-sd}(a), the average network cost for NDAP is much lower compared to the other algorithms when the same number of AEs are deployed (as the simulation terminates when any of the algorithms fail to deploy an AE in the group scenario) as, on average, it incurs 61\% and 38\% less cost compared to NVA and FFD, respectively. Moreover, for each algorithm, the cost increases with the increase of workload variations. This is due to  the fact that for higher variation in resource demands, the algorithms experience reduced scope in the data center for AE components placement as the feasibility domain is shrunk. As a consequence, feasible placements incur increasingly higher network cost with the increase of $sd$ parameter.

In the individual scenario, NDAP outperforms other algorithms in terms of the number of AEs deployed across various workload variations (Figure \ref{chap4-fig-perf-sd}(b)) by successfully placing on average 12\% and 15\% more AEs compared to NVA and FFD, respectively. Due to the random placement component, overall NVA performs better compared to FFD which is deterministic by nature. Another general pattern noticeable from the chart is that, all the algorithms deploy more AEs for lower values of $sd$. This is due the fact that for higher value of $sd$, resource demands of the AE components demonstrate higher variations and, as a consequence, resources of data center components get more fragmented during the AE placements and, thus, the utilization of those resources get reduced. 

\subsection{Scaling Network Distances}
\label{chap4-sec-scaling-net-dist}
This experiment varies the relative network distance between any two data center nodes by scaling the $DF$ parameter defined in section \ref{chap4-sec-data-center-setup}. As the definition implies, the inter-node network distance increases with $DF$ and such situation can arise due to higher delays in network switches or due to geographical distances. Initially, the $DF$ value is set to 2 and increased upto 16. Other parameters $N$, $meanCom$, $meanStr$, $meanVLBW$, and $sd$ are set to 1152, 0.3, 0.4, 0.35, and 0.5, respectively.

\begin{figure}[t]
\centering
\includegraphics[scale=0.6, trim=1cm 32.5cm 0cm 2cm]{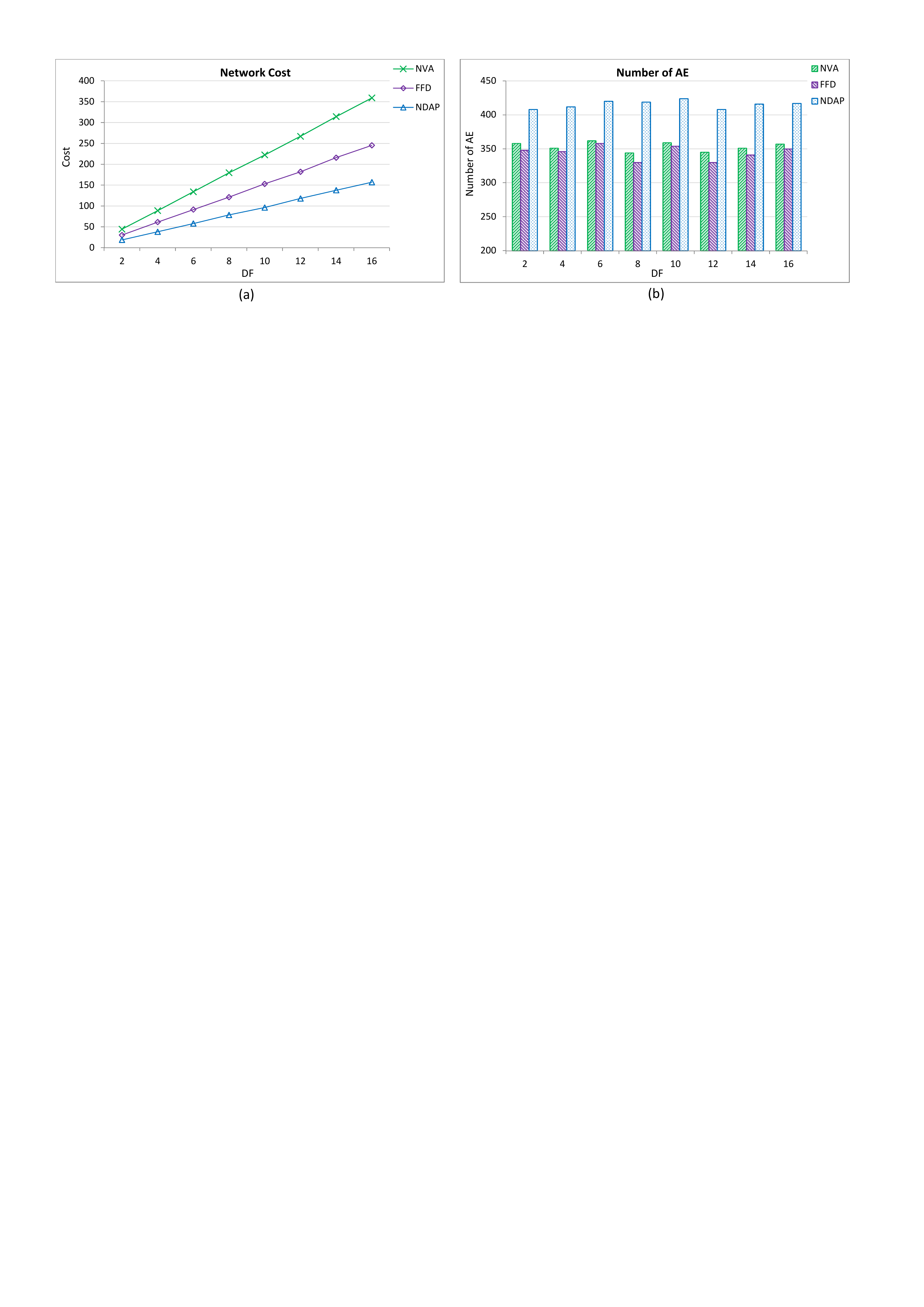}
\caption{Performance with increasing distance factor DF: (a) Network cost and (b) Number of AE deployed in DC.}
\label{chap4-fig-perf-DF}
\end{figure}

Since network distance directly contributes to the cost function, it is evident from Figure \ref{chap4-fig-perf-DF}(a) that the placement cost rises with the increase of the $DF$ parameter in a linear fashion for the group scenario. Nevertheless, the gradients for the different placement algorithms are not the same and the rise in cost for NDAP is much lower than other algorithms. 

Figure \ref{chap4-fig-perf-DF}(b) shows the average number of AEs deployed in data center for each $DF$ values for the individual scenario. Since network distance does not contribute to any of the resource capacities or demands (e.g., CPU or bandwidth), the number of AE deployment remains mostly unchanged with the scaling of $DF$. Nonetheless, through efficient placement, NDAP outpace other algorithms and successfully deploys 18\% and 21\% more AEs than NVA and FFD, respectively. 


\subsection{Network Utilization}
\label{chap4-sec-net-util}
This part of the experiment is conducted for the purpose of comparing the network utilization of the placement algorithms at the access, aggregation, and core switch levels of the data center network. This is done by stressing the network in two different scaling factors separately: the mean and the standard deviation of VLs bandwidth demand $meanVLBW$ and $sdVLBW$, respectively. In order to ensure that the computing and storage resource demands (of VMs and DBs, respectively) do not stress the computing and storage resource capacities (of the CNs and SNs, respectively), the $meanCom$ and $meanStr$ parameters are kept at a fixed small value of 0.05, and the standard deviation $sdComStr$ for both computing and storage resource demands is set to 0.1. As for the other parameters, $N$ and $DF$ are set to 1152 and 2, respectively.

\subsubsection{Scaling Mean Bandwidth Demand}
\label{chap4-sec-scale-vl-bw-dem}
Here the data center network is stressed using the group scenario where all the algorithms terminate if any of the algorithms fail to place an AE. Application of the group scenario for this experiment ensures that the total network loads imposed for each of the placement algorithms are same when any of the algorithms fail. Initially, the mean VL bandwidth demand $meanVLBW$ is set to 0.1 and raised upto 0.7, each time increasing it by 0.1. Standard deviation of VL bandwidth demand $sdVLBW$ is kept fixed at 0.3.

\begin{figure}[t]
\centering
\includegraphics[scale=0.55, trim=0cm 23.7cm 2.4cm 2cm]{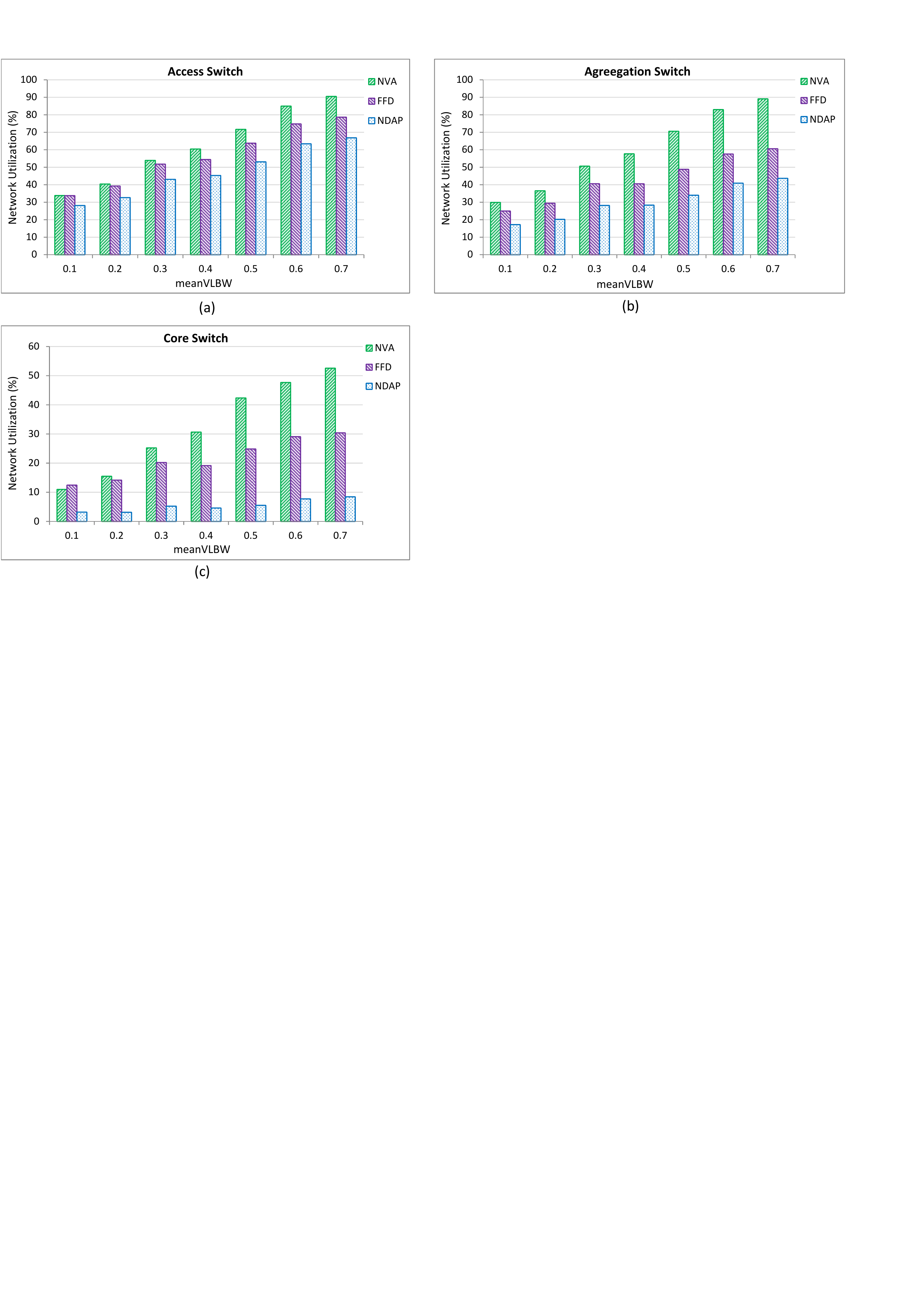}
\caption{Average network utilization with increasing mean VL bandwidth demand: (a) Access switch, (b) Aggregation switch, and (c) Core switch.}
\label{chap4-fig-perf-net-util-meanVLBW}
\end{figure}

Figure \ref{chap4-fig-perf-net-util-meanVLBW} shows the average network utilization of the access, aggregation, and core switches for different $meanVLBW$ values. It is evident from the charts that, for all the switch levels, NDAP incurs minimum average network utilization and compared to NVA and FFD, NDAP placements on average result in 24\% and 16\% less network usage for access layer, 49\% and 30\% less network usage for aggregation layer, and 83\% and 75\% less network usage for core layer. This represents the fact that NDAP localizes network traffic more efficiently compared to other algorithms and achieves incrementally higher network efficiency at access, aggregation, and core switch levels. Furthermore, as the figure demonstrates, the results reflect a similar trend of performance as the results of average network cost for placement algorithms presented in the previous subsections. This is reasonable since network cost is proportional to the distance and bandwidth of the VLs and while placing a VL, higher network distance indicates the use of higher layer switches. Thus, these results validate the proposed network cost model (Eq. \ref{chap4-eq-cost-1} \& \ref{chap4-eq-cost-2} ) in the sense that indeed it captures the network load perceived by the network switches. Also, it can be observed that the utilization for each switch increases with increasing $meanVLBW$. This is due to the fact that $meanVLBW$ contributes to the average amount of data transferred through the switches since $meanVLBW$ is used as the mean to generate the VLs bandwidth demands.

\subsubsection{Diversification of Bandwidth Demand}
\label{chap4-sec-div-bw-dem}
This experiment is similar to the above one, however, here the standard deviation of VLs bandwidth demands ($sdVLBW$) is scaled rather than the mean--- initially, $sdVLBW$ is set to 0.05 and gradually increased upto 0.5, each time raising it by 0.05. The mean VLs bandwidth demand $meanVLBW$ is set to 0.4. 

\begin{figure}[H]
\centering
\includegraphics[scale=0.55, trim=0cm 24.5cm 0.2cm 2cm]{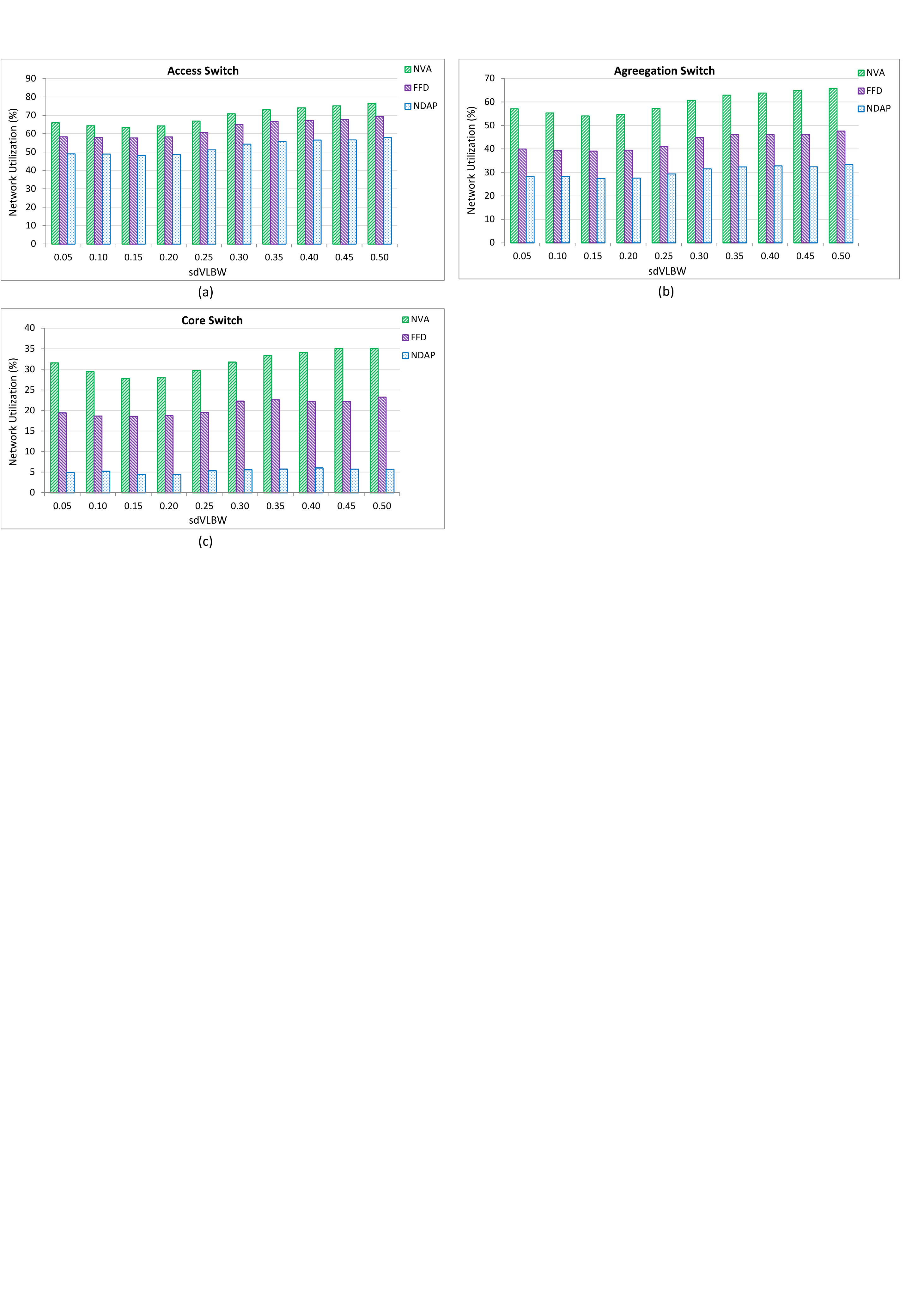}
\caption{Average network utilization with increasing standard deviation of VL bandwidth demand: (a) Access switch, (b) Aggregation switch, and (c) Core switch.}
\label{chap4-fig-perf-net-util-sdVLBW}
\end{figure}

The results of this experiment is shown in Figure \ref{chap4-fig-perf-net-util-sdVLBW}. The charts clearly demonstrate the supervisor performance of NDAP that causes minimum network usage across all switch levels and compared to NVA and FFD, it has on average 26\% and 16\% less network usage for access layer, 50\% and 30\% less network usage for aggregation layer, and 84\% and 75\% less network usage for core layer. Furthermore, the figure shows that the network utilizations for each algorithm at each layer across different $sdVLBW$ values do not fluctuate much. This is due to the fact that although the variation of VLs' bandwidth demand increases with increasing $sdVLBW$, the overall network load levels do not change much and, as a result, the average network loads perceived by the network switches at different layers differ in a small range. 

\subsection{NDAP Decision Time}
\label{chap4-sec-ndap-dec-time}
\begin{figure}[t]
\centering
\includegraphics[scale=0.57, trim=0cm 16.3cm 0cm 2cm]{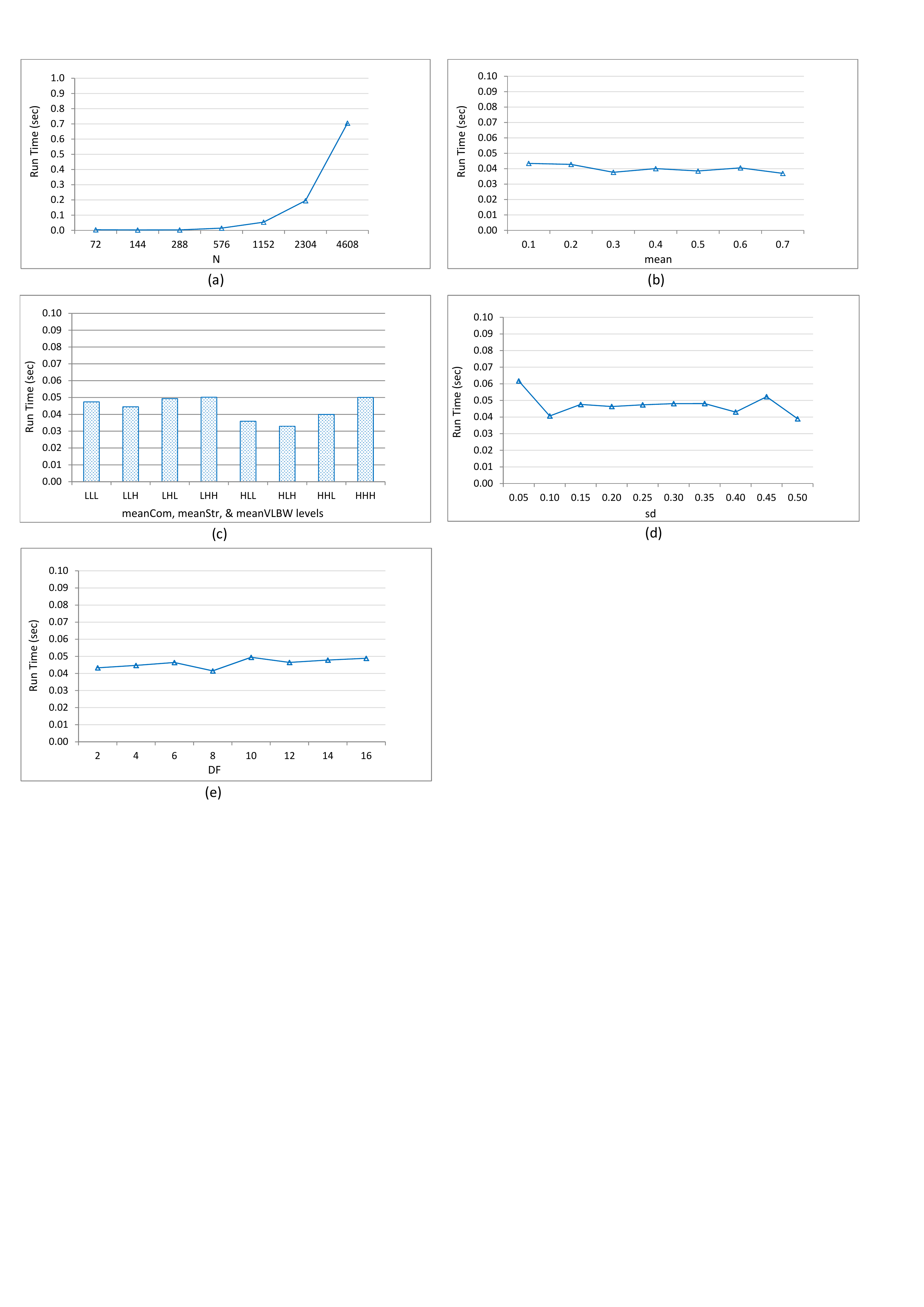}
\caption{NDAP's placement decision time while scaling (a) Data center size ($N$), (b) Homogeneous mean ($mean$), (c) Heterogeneous mean ($meanCom$, $meanStr$, $meanVLBW$), (d) Diversification of workload ($sd$), and (e) Distance factor ($DF$).}
\label{chap4-fig-perf-decision-time}
\end{figure}

In this part of the experiment, the time taken by NDAP for making AE placement decision is measured in order to assess the feasibility of using NDAP for real-time, on-demand placement scenarios. Figure \ref{chap4-fig-perf-decision-time} shows the average time needed by NDAP for computing AE placements in the individual scenario by scaling all the above mentioned scaling factors. For each of the scaling factors, other parameters are set similar to the corresponding preceding sections (Section \ref{chap4-sec-scaling-dc-size}-\ref{chap4-sec-scaling-net-dist}).

From Figure \ref{chap4-fig-perf-decision-time}(a), it can be seen that NDAP's run time increases non-linearly with increasing data center size (other parameters are set as in Section \ref{chap4-sec-scaling-dc-size}). It is evident from the figure that for small to medium data centers, NDAP's decision making time is in the range of a small fraction of a second, whereas for the largest data center simulated with $N=4608$ (i.e., several thousand servers), NDAP needs only about 0.7 second. Furthermore, it can be observed from Figure \ref{chap4-fig-perf-decision-time}(b)-(e) that NDAP's run time remains largely unaffected by other scaling factors and the run time is within the range of 0.03-0.06 second for $N=1152$. 
From the above results and discussion, it can be concluded that NDAP is suitable for on-demand AE placement scenarios, even for large data centers.

\color{black}

\section{Conclusions and Future Work}
\label{chap4-sec-con-future-work}
With the growing complexity of modern Internet applications, as well as size of data, network resource demands in large data centers, such as Clouds, are becoming increasingly complex. Rising bandwidth requirements among application components are causing rising pressure on the underlying communication infrastructure and, thus, make it a key area of performance bottleneck. This paper addressed the issue of network-focused, multi-component application placement in large data centers and formally defined it as an optimization problem. After presenting the constitutional components of the proposed network- and data-aware application placement approach, it presented NDAP, a greedy heuristic that performs simultaneous deployment of VMs and data components respecting computing, network, and storage resource requirements and capacity constraints with the goal of minimizing the network cost incurred due to the placement decision. Moreover, a detailed analysis on the computational complexity for each of the compared algorithms in terms of run-time and memory overhead is presented. It is revealed from the analysis that the proposed NDAP algorithm has quadratic time complexity, which is slightly higher than those of the compared algorithms (linear and linearithmic), and the memory overheads are same for each of the algorithms. Furthermore, performance of the proposed NDAP algorithm is compared with related works through extensive simulation-based experiments and the results demonstrate superior performance of NDAP over competitor approaches across multiple performance metrics.

The proposed NDAP algorithm is developed as a generic application placement heuristic, which is very application-unaware, so that it can be used for a wide range of multi-tier or composite applications. The effectiveness of NDAP is validated in minimizing overall communications overhead for two different representative application environments: multi-tier and scientific (Montage) workflows. A further optimization of application placement, considering application-aware communication models and specific structure for application components, can be a potential future work. Moreover, the proposed NDAP strategy is flexible enough to be extended in future to accommodate such special cases.

Widespread use of virtualization technologies, high speed communication, increased size of data and data centers, and, above all, the broad spectrum of modern applications are opening new research challenges in network resource optimization. Appropriate combination and coordination of the online and offline VM placement and migration techniques with the goal of efficient network bandwidth management is one of the key areas for future research. 

Furthermore, periodic and threshold-based reconfiguration of application virtual components using VM migration and reallocation with focus on network traffic localization can be an effective strategy for data center resource optimization. Such online or real-time optimization techniques must employ appropriate resource demand forecasting for both computing (e.g., CPU utilization) and network resources (e.g., bandwidth usage). In addition, VM migration and reconfiguration overhead can have significant impact on the performance of the hosted applications, as well as on the underlying communication substrate, consequently, questioning the scalability and viability of the offline strategies that often consider simplistic measures for VM migration overheads. Incorporation of reconfiguration overhead estimation with offline optimization techniques focusing on multiple objectives will produce more pragmatic VM migration schemes trading off between resource usage optimization and incurred overhead.




\bibliographystyle{elsarticle-harv} 
\bibliography{D:/Sunny/Dropbox/monash/bibtex-references/references-all}






\end{document}